\documentclass[runningheads]{llncs}
\usepackage{graphicx}
\PassOptionsToPackage{hyphens}{url}\usepackage{hyperref}
\usepackage{bibnames}
\usepackage{array, booktabs} 
\usepackage{multirow, xspace}
\usepackage{xcolor}
\usepackage{scrextend}
\usepackage[T1]{fontenc}
\usepackage{lscape}
\usepackage{longtable}
\newcolumntype{C}[1]{>{\centering\arraybackslash}p{#1}}
\newcolumntype{R}[1]{>{\raggedleft\arraybackslash}p{#1}}
\newcolumntype{L}[1]{>{\raggedright\arraybackslash}p{#1}}

\usepackage[ruled]{algorithm2e} 

\SetAlFnt{\small}
\SetAlCapFnt{\small}
\SetAlCapNameFnt{\small}
\SetAlCapHSkip{0pt}
\IncMargin{-\parindent}

\newcommand{\eg}{\emph{e.g.}\xspace}
\newcommand{\ie}{\emph{i.e.}\xspace}

\begin{document}
\title{An Extended Pattern Collection \\for Blockchain-based Applications}
%
%
\author{Xiwei Xu\inst{1,2} \and
Cesare Pautasso\inst{3} \and \\
Sin Kuang Lo\inst{1,2} \and
Liming Zhu\inst{1,2} \and
Qinghua Lu\inst{1,2} \and
Ingo Weber\inst{4,5}}
\authorrunning{X. Xu et al.}

%
\institute{CSIRO'Data61, Sydney, Australia\\ 
\email{\{firstname\}.\{secondname\}@data61.csiro.au}\\
\and
School of Computer Science and Engineering, University of New South Wales, Sydney, Australia\\
\and
University of Lugano,  Lugano, Switzerland\\
\email{c.pautasso@ieee.org}\\
\and
Technical University of Munich, School of CIT, Munich, Germany\\
\email{firstname.lastname@tum.de}
\and
Fraunhofer-Gesellschaft, Munich, Germany\\
}
\maketitle              
\begin{abstract}
Blockchain is an emerging technology that enables new forms of decentralized software architectures, where distributed components can reach agreements on shared system states without trusting a central integration point. Blockchain provides a shared infrastructure to execute programs, called smart contracts, and to store data. Since blockchain technologies are at an early stage, there is a lack of a systematically organized knowledge providing a holistic view on designing software systems that use blockchain. We view blockchain as a component of a bigger software system, which requires patterns for using blockchain in the design of the software architecture. In this paper, we collect a list of patterns for blockchain-based applications. The pattern collection is categorized into five categories, including interaction with external world patterns, data management patterns, security patterns, structural patterns of contracts, and user interaction patterns. Some patterns are designed considering the nature of blockchain and how blockchains can be specifically introduced within real-world applications. Others are variants of existing design patterns applied in the context of blockchain-based applications and smart contracts. 

\end{abstract}

\section{Introduction}

Blockchain is the technology behind Bitcoin~\cite{Satoshi:bitcoin}, which is a digital currency based on a peer-to-peer network and cryptographic techniques. The blockchain provides immutable, append-only, shared data storage, which only allows inserting transactions without updating or deleting any existing ones, thus preventing any tampering or revision of previously stored data on blockchain as long as the majority of the network peers do not agree to allow such revision. The blockchain enables decentralization as new forms of distributed software architectures, where components can reach agreements on the historical log of shared states for decentralized and transactional data sharing, across a large network of untrusted participants without relying on a central integration point.

Financial transactions are the first, but far from the only use case being investigated for blockchain. Many start-ups, enterprises, and governments~\cite{ukreport} are exploring blockchain-based applications in areas as diverse as supply chain, electronic health records, voting, energy supply, ownership management, and protecting critical civil infrastructure. Despite of the wide array of interest in blockchain technology, there is a lack of a systematic and holistic view when applying blockchain in the design of software applications.

Previous work has characterized blockchain from a software architecture perspective as a software connector~\cite{sherry2016} that provides a shared infrastructure for storing data and running programs (known as \emph{smart contracts}). Blockchain guarantees unique properties including immutability, non-repudiation, data integrity, transparency, and equal rights. It also has two main limitations, namely, lack of privacy and poor performance~\cite{sherry2016}. The taxonomy presented in~\cite{Sherry:ICSA2017} discusses such properties for different types and configurations of blockchain technology. To better leverage the positive properties of blockchain and avoid or reduce the impact of its limitations, more architectural guidance on blockchain-based applications is needed.

In this paper, we present a set of patterns for the design of blockchain-based applications. In software engineering, a design pattern is a reusable design solution to a problem that commonly occurs within a given context during software design~\cite{Beck1987}. A design pattern defines constraints that restrict the roles of architectural elements (processing, connectors and data) and the interaction among those elements. Adopting a design pattern causes trade-offs among quality attributes. Our pattern collection includes four patterns about interaction between blockchain and the external world, four data management patterns, four security patterns, five structural patterns of smart contract and two user interaction patterns. The pattern collection provides an architectural guidance for developers to build applications on blockchain. 

The remainder of the paper is organized as follows. Section~\ref{sec:background} presents a background of blockchain and smart contracts. Section~\ref{sec:patternCollection} gives an overview of the pattern collection, followed by detailed patterns discussed from Section~\ref{sec:interactionPatttern} to Section~\ref{sec:deploymentPatttern}. Related work on blockchain-based applications and design patterns is discussed in Section~\ref{sec:relatedwork}. Section~\ref{sec:conclusions} concludes the paper and outlines the future work.

\section{Background}
\label{sec:background}

\subsection{Blockchain}

Blockchain is a data structure of an ordered list of blocks, where every block "chained" back to the previous block through containing a hash of a presentation of the previous block. Every block on blockchain contains a list of transactions (possibly empty). A transaction is a data package that stores information for money transfer, like sender, receiver, and monetary value, or the (compiled) code of smart contracts, or parameters of function calls of smart contracts. Due to the security properties of hash function, the historical transactions on blockchain can not be deleted or altered without invalidating the chain of hashes. In addition to the design of the data structure, there are computational constraints and consensus protocols applied to the creation of blocks. All together, blockchain can in practice prevent revision and tampering of the information on blockchain. 

When using a blockchain, one design decision is the deployment, \ie~, whether to use a public blockchain, consortium/community blockchain or private blockchain~\cite{Sherry:ICSA2017}. Most cryptocurrencies use public blockchains, which can be accessed by anyone on the Internet. 
Using a public blockchain results in better information transparency and auditability, but sacrifices performance and has a different cost model compared with a conventional data storage. It costs monetary value to store data or execute code on a public blockchain. In a public blockchain, data privacy relies on encryption or cryptographic hashes. A consortium blockchain is used across multiple organizations. The consensus process in a consortium blockchain is controlled by a set of pre-authorised nodes. The right to read the blockchain may be public or may be restricted to specific participants. In a private blockchain network, write permission is kept within one organization, although this may include multiple divisions of a single organization. 

\subsubsection{Properties}
\label{sec:properties}

The data contained in a transaction on blockckchain is seen as \textit{immutable} in practice. The chain of immutable cryptographically-signed historical transactions provides \textit{non-repudiation} of the stored data. Cryptographic techniques used by blockchain support data \textit{integrity}, the public access provides data \textit{transparency}, and \textit{equal rights} allows every participant to have the same ability to manipulate the data on blockchain. Such rights can be weighted by the computational power (\emph{Proof-of-work}) or stake (\emph{Proof-of-stake}) owned by a node. \textit{Trust} of the blockchain is built based on the interactions between nodes within the blockchain network. The participants of a blockchain network rely on the design of blockchain, the cryptographic techniques used by blockchain and the blockchain network itself rather than relying on trusted third-party to facilitate transactions. 

\subsubsection{Limitations}
\label{sec:limitations}

\textit{Data privacy} and \textit{scalability} are the main two limitations of public blockchains. Data privacy on public blockchain is limited because there is no privileged user, and every participant can join the network to access all the information on blockchain and validate new transactions. There are scalability limits on (i) the size of the data included into a transaction, (ii) the transaction processing rate, and (iii) the latency of data transmission and commits. Latency between submitting a transaction and it being committed on a blockchain is affected by the consensus protocol. This is around 1 hour (10-minute block interval with time for inclusion and 5-block confirmation) on Bitcoin, and around 3 minutes (14-second block interval with 11 confirmation blocks) on Ethereum\footnote{\url{https://www.ethereum.org/}}. Times in practice can be even longer~\cite{SRDS2017}. The number of transactions included in each block is also limited by the bandwidth of nodes participating in the network (for Bitcoin the current bandwidth per block is 1MB)~\cite{blockstack2016}. Ethereum applies a so-called \textit{gas} limit to blocks (\textit{gas} is the internal pricing unit for executing a transaction or storing data), which limits the number and complexity of transactions that can fit into a block. 

\subsubsection{Blockchain as a Software Component}

When used in a large software system, blockchain can be viewed as a software component~\cite{sherry2016}. In such software system, blockchain is responsible for storing and sharing data, and executing smart contracts. Due to the limitations of privacy and performance, there might be off-chain auxiliary databases used in the system. For example, private or large sized data can be stored in an internal database. There is normally a API layer between the data storage layer and the applications using the blockchain, which is same as with conventional technology. When blockchain interacts with other off-chain components, an ``oracle''~\cite{lo2020reliability} is needed to bridge blockchain with the external world.

\subsection{Smart Contract}
\label{sec:smartcontract}

The first generation of blockchains, like Bitcoin, provides a public ledger to store cryptographically-signed financial transactions~\cite{Swan:blockchain}. 
There is very limited capability to support programmable transactions, and only very small pieces of auxiliary data could be embedded in the transactions to serve other purposes, such as representing other digital assets or physical assets.

The second generation of blockchains provides a general-purpose programmable infrastructure with a public ledger that records the computational results. Programs, known as \emph{smart contracts}~\cite{Omohundro:2014}, can be deployed and run on a blockchain. Smart contracts can express triggers, conditions and business logic~\cite{Weber:BPM2016} to enable more complex programmable transactions. The signature of the transaction initiator authorizes the data payload of a transaction or the creation or execution of a smart contract. A common simple example of a smart contract-enabled service is escrow, which can hold funds until the obligations defined in the smart contract have been fulfilled. Smart contracts are pure functions by design, which cannot access the state of external systems directly. 

\subsubsection*{Smart Contract Languages}

\textit{Script} used by Bitcoin is a simple stack-based scripting language\footnote{\url{https://en.bitcoin.it/wiki/Script}}, which is intentionally designed not to be Turing-complete. Script provides the flexibility to define conditions required to spend the Bitcoin associated with the transactions, for example, requiring multiple private keys to authorize the payment. Ethereum is currently the most widely-used blockchain that supports general-purpose (Turing-complete) smart contracts. The primary smart contract language used on Ethereum blockchain is \textit{Solidity}\footnote{\url{https://solidity.readthedocs.io/}}. DigitalAsset\footnote{\url{http://www.digitalasset.com/}} proposed DAML\footnote{\url{https://daml.com/}}as a domain specific smart contract language for financial institutes. Smart contracts running on Hyperledger Fabric\footnote{\url{https://www.hyperledger.org/use/fabric}} are called Chaincode, which can be written in any programming language and executed in containers inside the fabric context layer.


\section{Overview of Blockchain-based Application Patterns}
\label{sec:patternCollection}

\begin{figure}[t]
\begin{center}
\includegraphics[width=1\columnwidth]{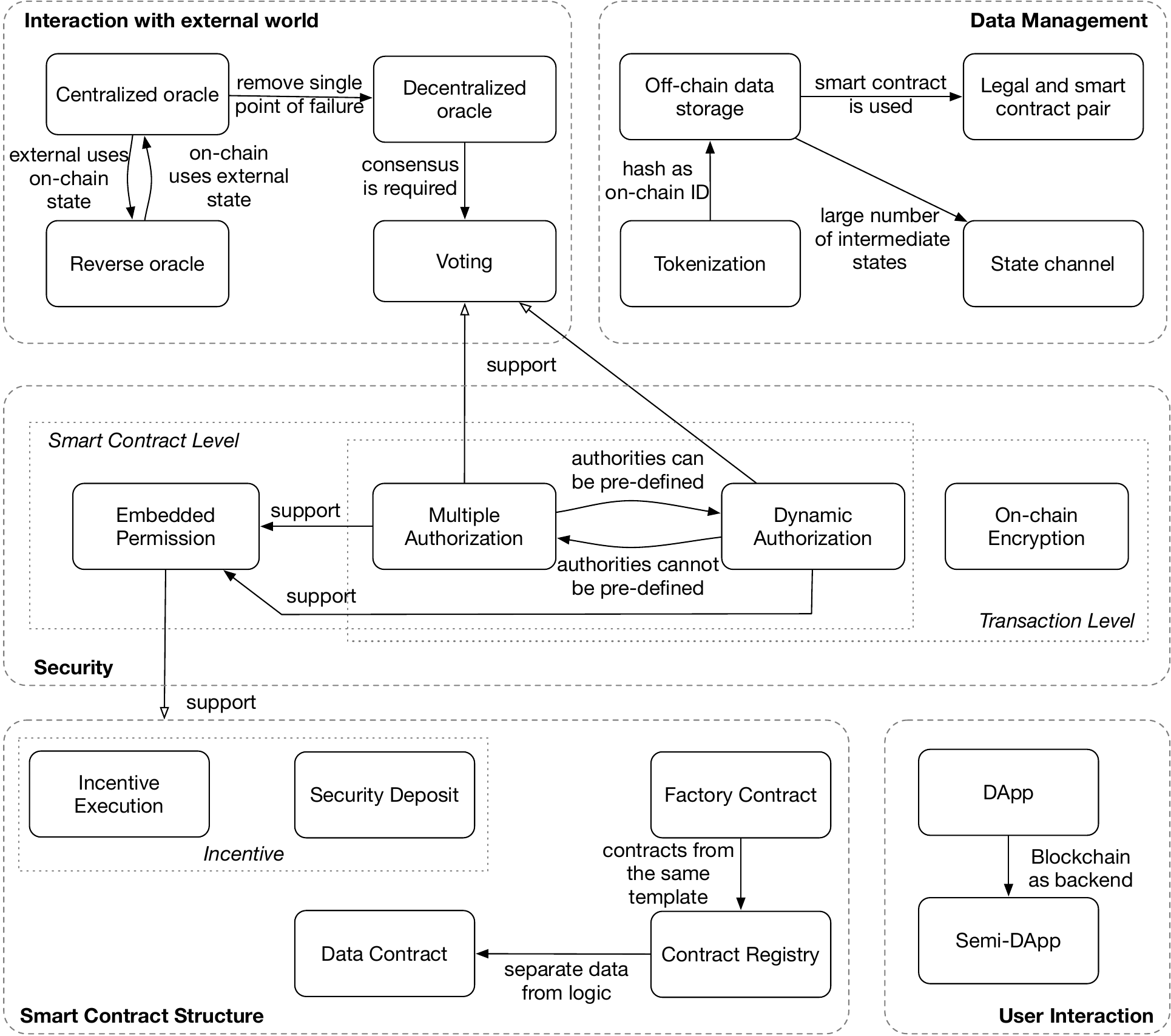}
\caption{Blockchain as a component within a software architecture}\label{fig:overview}
\end{center}
\end{figure}

In this section, we discuss the overview of the blockchain-based application pattern collection, which currently includes nineteen design patterns that shape the architectural elements and their interactions in blockchain-based applications. Figure.~\ref{fig:overview} gives an overview of these patterns as well as their mutual relationships. The arrows in Figure.~\ref{fig:overview} illustrate how to navigate through the pattern collection during a design process. Applying the patterns to an blockchain-based application can better align it with the unique properties provided by blockchain, avoid its limitations, and achieve other quality attributes.

The patterns about interaction between blockchain and the external world describe different ways for blockchain to communicate data with the external world, including \textit{Centralized oracle} (Section~\ref{sec:centralOracle}), \textit{Decentralized oracle} (Section~\ref{sec:decentralOracle}), \textit{Reverse oracle} (Section~\ref{sec:reverseverifier}) and \textit{Voting} (Section~\ref{sec:voting}). The four data management patterns are about managing data on and off blockchain, including \textit{Tokenisation} (Section~\ref{sec:tokenisation}), \textit{Off-chain data storage} (Section~\ref{sec:hash}), \textit{State channel} (Section~\ref{sec:statechannel}) and \textit{Legal and smart contract pair} (Section~\ref{sec:legal}). The four security patterns concern the security aspect of the blockchain-based applications. \textit{Multiple authorization} (Section~\ref{sec:authority}) and \textit{Dynamic authorization} (Section~\ref{sec:dynamic}) are aimed at adding dynamism to authorization of transactions and smart contracts. \textit{Embedded permission} (Section~\ref{sec:permission}) aims to improve security of smart contracts, and \textit{On-chain Encryption} (Section~\ref{sec:encrypted}) can be used to improve security of on-chain data.

The five contract structural patterns define the dependencies among smart contracts and behaviour of smart contract. Smart contracts on blockchain are immutable. Upgrading a smart contract to a new version is a challenge which hinders the evolution of blockchain-based applications. \textit{Contract registry} (Section~\ref{sec:registry}), \textit{Factory contract} (Section~\ref{sec:factory}) and \textit{Data contract} (Section~\ref{sec:separate}) can used together to improve upgradability of smart contracts. \textit{Incentive execution} (Section~\ref{sec:incentiveexecution}) and \textit{Security deposit} (Section~\ref{sec:securitydeposit}) provide incentive mechanism for execution and maintenance of smart contracts. The two user interaction patterns summarise different ways for users to interact with DApps, including \textit{DApp} (Section~\ref{sec:DApp}) and \textit{Semi-DApp} (Section~\ref{sec:SemiDApp}).


In this paper we follow the extended pattern form from~\cite{meszaros1998pattern}, which includes the name of the pattern, a short summary, the context, the problem statement, an explicit discussion of the forces which make the problem difficult, the solution, its consequences, and some examples of real-world known uses of the pattern. 
Forces are identified with the corresponding quality attribute, as sometimes the solution will propose a trade-off between them. Regarding the consequences, we distinguish the benefits and drawbacks. Finally, we discuss features only applicable to a certain deployment of blockchain, such as monetary cost of data storage and code execution.

\section{Interaction with External World Patterns}
\label{sec:interactionPatttern}

As a component of a big software system, blockchain needs to communicate data with other components within the software system. This section discuss four patterns applicable to the interaction between blockchain and external world.


\subsection{\textbf{Pattern 1: Centralized Oracle}}
\label{sec:centralOracle}

\begin{figure}[t]
\begin{center}
\includegraphics[scale=0.9]{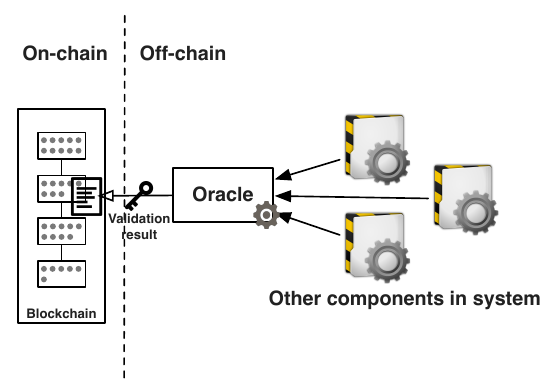}
\caption{Centralized Oracle Pattern}\label{fig:centralOracle}
\end{center}
\end{figure}

\noindent \textbf{Summary:} Introduce the state of external systems into the closed blockchain execution environment through a single centralized connector (called \textit{oracle}). Fig.~\ref{fig:centralOracle} is a graphical representation of the pattern.

\vspace{0.5em}\noindent \textbf{Context:} 
From the software architecture perspective, blockchain can be viewed as a component or connector within a large software system~\cite{sherry2016}. In the case where blockchain is used as a distributed database for more general purposes other than financial services, the applications built on blockchain might need to interact with other external systems. Thus, the validation of transactions on blockchain might depend on states of external systems. 

\vspace{0.5em}\noindent \textbf{Problem:} 
Smart contracts running on blockchain are pure functions by design. The execution environment of smart contract is self-contained. It can only access information present in the data and transactions on the blockchain.The state of external systems are not directly accessible to smart contracts. 
How can function calls in smart contracts be enabled to access the state of the external world from within smart contracts? 

\vspace{0.5em}\noindent \textbf{Forces:} 
The problem requires to balance the following forces:
\begin{itemize}
  \item \textit{Closed environment.} Blockchain is a secure, self-contained environment, which is isolated from external systems. Smart contracts on blockchain cannot read the states of the external systems.
  \item \textit{Connectivity.} In addition to the data found on the blockchain, general-purpose applications might require information from external systems. For example, a parcel tracking application needs context information like geo-location information, a gambling application might need weather data from a Web API\footnote{\url{https://openweathermap.org/api}}.
\end{itemize}

\vspace{0.5em}\noindent \textbf{Solution:} 
To connect the closed execution environment of blockchain with the external world, \textit{oracle} is introduced to assist in evaluating conditions that cannot be expressed in a smart contract running within the blockchain environment. If the information flows from blockchain to external world, Reverse Oracle (Section~\ref{sec:reverseverifier}) should be used. An centralized oracle is a trusted third party that provides the smart contracts with information about the external world. When validation of a transaction depends on external state, the oracle is requested to check the external state and inject the result to the blockchain in a transaction signed using its own key pair. The validators (\emph{miner}) take the result provided by the oracle into account when validating the transaction. From the perspective of validator, by introducing the oracle, the validation of transactions is based on the authentication of the oracle (through digital signature) rather than the external state because what provided by oracle is trusted by the validator. From the perspective of the validator, the data injected from oracle is no different from data provided (through embedding into a transaction or as a variable value) by other users. What validators can do is to ensure the data integrity by checking the digital signature of the sender and executing the smart contract based on the input data, but they cannot check the originality or correctness of the input data from external world. More technical details of implementing the pattern could be found in ~\cite{2020-Muehlberger-BPM-BC}. 


\vspace{0.5em}\noindent \textbf{Consequences:} 

Benefits:
\begin{itemize}
  \item \textit{Connectivity.} The closed execution environment of blockchain is connected with external world through a centralized oracle. The applications based on blockchain can access external states through the oracle and use the external states to validate transactions.
  \item \textit{Efficiency.} Centralized oracle is more efficient in terms of monetary and time cost. It is easier to manage compared with Decentralized Oracle (Section~\ref{sec:decentralOracle})
\end{itemize}

Drawbacks: 
\begin{itemize}
   \item \textit{Trust.} Using a centralized oracle introduces a trusted third party into the system. The oracle selected to verify the external state needs to be trusted by all the participants involved in relevant transactions.  
   \item \textit{Validity.} The external states injected into the transactions can not be fully validated by miners. Thus, when miners validate the transaction including external state, they rely on the oracle to check the validity of the information from external world. 
   \item \textit{Long-term availability and validity.}
   It could happen that while transactions are immutable, the external state used to validate them may change after the transactions were originally appended to the blockchain. 
   \item \textit{Single point of failure.} A centralized oracle introduces a single trusted element, whose unavailability or failure may prevent the blockchain from successfully completing the transaction verification process.
\end{itemize}

\vspace{0.5em}\noindent \textbf{Related patterns:} 
\begin{itemize}
    \item \textit{Decentralized Oracle} (Section~\ref{sec:centralOracle}) can be used to avoid the single point of failure introduced by a centralized oracle.
    \item \textit{Reverse Oracle} (Section~\ref{sec:reverseverifier}) can be used when the information flows from blockchain to external components.
\end{itemize}

\vspace{0.5em}\noindent \textbf{Known uses:}
\begin{itemize}
  \item \emph{Oracle} in Bitcoin is an instance of this pattern \footnote{\url{https://en.bitcoin.it/wiki/Contract\#Example_4:_Using_external_state}}. Oracle is a server outside the Bitcoin blockchain network, which can evaluate user-defined expressions based on the external state.
  \item \emph{Provable}\footnote{\label{note}\url{https://provable.xyz/}} is an oracle service provider, which utilises trusted hardware to directly fetch information from external trusted execution environment (TEE). Provable introduces three different proofs for fetching data from external data sources, namely, TLS-Notary, Ledger proof and Android proof.
  \item \emph{Corda}\footnote{\label{corda}\url{https://www.corda.net/}} has a centralized oracle mechanism embedded in its platform. The oracle mechanism uses Intel Software Guard Extensions (SGX) for hardware attestation to prevent unauthorised access outside of the SGX environment.
\end{itemize}


\subsection{\textbf{Pattern 2: Decentralized Oracles}}
\label{sec:decentralOracle}

\begin{figure}[t]
\begin{center}
\includegraphics[scale=0.9]{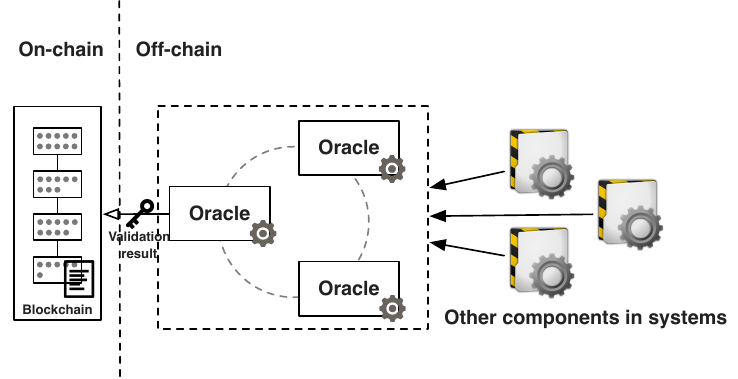}
\caption{Decentralized Oracle Pattern}\label{fig:decentraloracle}
\end{center}
\end{figure}

\noindent \textbf{Summary:} Introduce the state of external systems into the closed blockchain execution environment through a cluster of connectors (called \textit{oracles}). Fig.~\ref{fig:decentraloracle} is a graphical representation of the pattern.

\vspace{0.5em}\noindent \textbf{Context:} 
In the case where blockchain is used as a distributed database for more general purposes other than financial services, the applications built on blockchain might need to interact with other external systems. A centralized \textit{oracle} (Section~\ref{sec:centralOracle}) can be applied to inject the states of external systems into blockchain.

\vspace{0.5em}\noindent \textbf{Problem:} 
A centralized oracle introduces a single trusted third party into the system, which might becomes a single point of failure of the whole software system.


\vspace{0.5em}\noindent \textbf{Forces:} 
The problem requires to balance the following forces:
\begin{itemize}
 \item \textit{Reliability and Availability.} A centralized oracle becomes a single point of failure from an architecture perspective. In the case that the status injected into the blockchain is a faulty status, the whole system might behave inaccurately. In the case that the oracle is unable to inject any state to the system, the whole system might be stuck and unavailable depending on how critical the state is. 
 \item \textit{Cost.} The cost of data retrieval from external world is proportional to the number of oracles.  
\end{itemize}

\vspace{0.5em}\noindent \textbf{Solution:} 
To improve trustworthiness of the oracle, a decentralized oracle mechanism is introduced, which is based on multiple oracles. These oracles can get data from one data source or multiple independent data sources. 
\textit{Voting} (Section~\ref{sec:voting}) can be applied to decentralized oracle to reach a consensus on the status to be injected into the blockchain. 

\vspace{0.5em}\noindent \textbf{Consequences:} 

Benefits:
\begin{itemize}
  \item \textit{Reliability and Availability.} By having multiple oracles retrieving data from external world, the risk of validating transactions based on faulty external data is reduced from a single centralized oracle. Acquiring data by multiple oracles also improves the reliability and the confidence of the final accepted value. 
\end{itemize}

Drawbacks: 
\begin{itemize}
   \item \textit{Trust.} Although using decentralized oracle avoids the single-point-of-trust, it still introduces trusted third parties into the system. All the oracles that verify the external state needs to be trusted by all the participants involved in relevant transactions. So trust needs to be extended from a single entity to a cluster. 
  \item \textit{Cost.} Cost of using a piece of data from external world increases with the number of oracles being used.
  \item \textit{Time.} It might take longer time for multiple oracles to get the required information and reach a consensus over the final result. If all oracles need to agree, the time is bound by the slowest oracle. If a subset of the oracles is sufficient, then the slowest oracles can be ignored, thus potentially speeding up the process.
  \item \textit{Uncertainty.} Compared with centralized oracle, decentralized oracle introduces more uncertainties. When human is involved, the value what majority vote becomes the result, which might be different from the \textit{truth} in the physical world. 
\end{itemize}

\vspace{0.5em}\noindent \textbf{Related patterns:} 
\begin{itemize}
    \item \textit{Centralized Oracle} (Section~\ref{sec:centralOracle}) can be used if a single oracle is trusted by all the participants involved into a transaction.
    \item \textit{Reverse Oracle} (Section~\ref{sec:reverseverifier}) can be used when the information flows from blockchain to external componnets.
    \item \textit{Voting} (Section~\ref{sec:voting})  (Section~\ref{sec:authority}) can be applied with decentralized oracles to achieve consensus among multiple oracles.
    
\end{itemize}

\vspace{0.5em}\noindent \textbf{Known uses:}
\begin{itemize}
  \item Orisi\footnote{\url{http://orisi.org/}} on Bitcoin maintains a set of independent oracles. Orisi allows the participants involved in a transaction to select a set of oracles and define the value of M (number of oracles) before initiating a conditional transaction.
  \item Gnosis\footnote{\label{gnosis}\url{https://gnosis.io/}} is a decentralized prediction market that allows users to choose any oracle they trust, such as another user or a web service, \eg~, for weather forecasts. A human oracle is also called \textit{arbitrator}, who is trusted by the interacting participants to resolve disputes or check external state. 
  \item Augur\footnote{\label{augur}\url{https://www.augur.net/}} is another prediction market that leverages the capability of human oracles to do prediction and resolve disputes.
\end{itemize}


\subsection{ \textbf{Pattern 3: Voting}}
\label{sec:voting}

\begin{figure}[t]
\begin{center}
\includegraphics[scale=0.9]{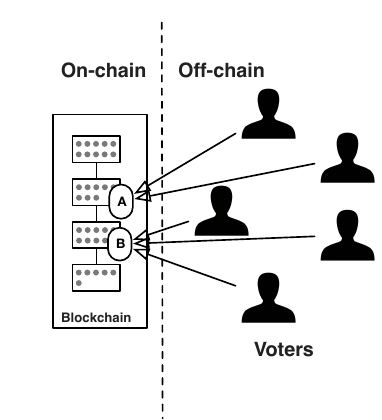}
\caption{Voting Pattern}\label{fig:voting}
\end{center}
\end{figure}

\noindent \textbf{Summary:} To achieve an agreement on a state proposed on blockchain, anyone with blockchain account can propose tentative new state or vote for a proposed state by staking their tokens until a consensus is achieve. Fig.~\ref{fig:voting} is a graphical representation of the pattern.

\vspace{0.5em}\noindent \textbf{Context:} 
The public access of blockchain provides \textit{equal rights} that allow~\cite{Sherry:ICSA2017} every participant the same ability to access and manipulate the blockchain. This property of blockchain enables a way for blockchain users to make decisions together and achieve consensus on the result. During the consensus process, everyone has equal right to participate in decision making. In the context of decentralized oracle, especially the one using human oracles, blockchain users use different sources to report the result. They may have different preferences. 

\vspace{0.5em}\noindent \textbf{Problem:} 
During a process to achieve an agreement on certain state on blockchain, what if the state proposed by an blockchain account (oracle or human) is disputed?

\vspace{0.5em}\noindent \textbf{Forces:} The problem requires to balance the following forces:
\begin{itemize}
  \item \textit{Fairness.} Every participant in blockchain network has a equal right to access and manipulate the blockchain. Each participant's vote should have the same weight as the others. 
  \item \textit{Consensus.} Multiple participants in different opinions need to reach an agreement to make decision. Participants also need to agree on which of the many paths leading to an outcome based on their preferences (e.g., simple or qualified majority) is taken.
  \item \textit{Transparency/Auditability.} The voting process should be deterministic and auditable so that the outcome can be reproduced from the same input (which should not disappear after the vote).
\end{itemize}

\vspace{0.5em}\noindent \textbf{Solution:} 

Voting is a mechanism commonly used by a group of participants to make a collective decision when the state originally proposed by participant is disputed. Anyone with a blockchain account that does not agree with the state can propose another state as tentative answer. To make a decision, every participant vote through sending transaction through her/his blockchain account. The voting transaction is signed by the private key of the participant, which represents the right of the participant to make decision. Such right can be weighted by the resource owned by the participant, like the application-specific tokens. Normally, majority rule is used to select the alternative which has the most votes (or with heavier weight in terms of stake) among all alternatives. 

One possible extension to the solution is to support secret voting by leveraging digital signatures. The voters can encrypt their choice when casting a vote, which can be decrypted using the corresponding public key when the votes are being counted. However, by linking the voting transaction with other transactions from and to the account, the anonymity of the user might be compromised. A more privacy-preserving way for the voter is to create a new account for voting only. It is debatable whether blockchain is a suitable technique to solve on-line voting security due to sociological issues that are outside of technical environment\footnote{\url{https://theconversation.com/blockchains-wont-fix-internet-voting-security-and-could-make-it-worse-104830}}.

\vspace{0.5em}\noindent \textbf{Consequences:} 

Benefits:
\begin{itemize}
  \item \textit{Equality.} Voting method allows the participants to use their right to participate decision making. 
  \item \textit{Consensus.} Multiple participants with different preferences can reach a consensus through voting.
\end{itemize}

Drawbacks: 
\begin{itemize}
  \item \textit{Duplication} The vote is associated with the blockchain account. Smart contract can help to avoid duplicated votes from the same blockchain account, for example by counting only the most recent vote. But since blockchain is pseudonymous, every participant can own multiple blockchain addresses to gain additional voting power, which is similar to the Sybil attack at network layer.  
  \item \textit{Time.} Voting may take a long time due to long voting/dispute time windows
\end{itemize}

\vspace{0.5em}\noindent \textbf{Related patterns:}
\begin{itemize}
    \item \textit{Decentralized Oracles} (Section~\ref{sec:decentralOracle}) works with voting pattern to achieve consensus on the answer reported to blockchain.
   \item \textit{Security Deposit} (Section~\ref{sec:securitydeposit}) provides a mechanism for participants to weight their vote using their stake. 
    \item \textit{Multiple Authorization} (Section~\ref{sec:authority}) is one on-chain mechanism to enable voting. 
    \item \textit{Dynamic Authorization} (Section~\ref{sec:authority}) is one off-chain mechanism to enable voting. 
\end{itemize}

\vspace{0.5em}\noindent \textbf{Known uses:}
\begin{itemize}
  \item Voting mechanism is used in DAOs (Decentralised Autonomous Organisations)\footnote{\url{https://www.ethereum.org/dao}}.
  \item In Gnosis\textsuperscript{\ref{gnosis}} prediction market, a voting mechanism is used if someone challenges the reported outcome. This voting mechanism allows users to vote on what the correct outcome was by betting Ether on that outcome. 
  \item In Augur\textsuperscript{\ref{augur}} prediction market, a similar voting mechanism is used to resolve disputes on the outcome reported by oracles.
\end{itemize}


\subsection{ \textbf{Pattern 4: Reverse Oracle}} 
\label{sec:reverseverifier}

\begin{figure}[t]
\begin{center}
\includegraphics[scale=0.9]{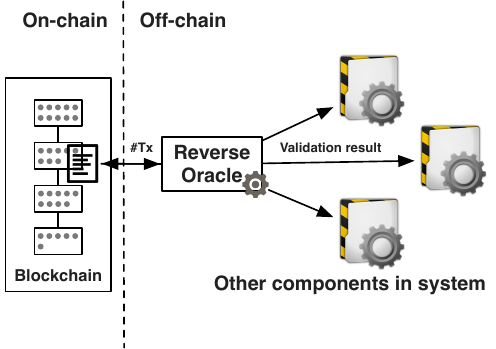}
\caption{Reverse Oracle Pattern}\label{fig:reverseverifier}
\end{center}
\end{figure}

\noindent \textbf{Summary:} The reverse oracle of an existing system relies on smart contracts running on blockchain to validate requested data and check required status. Fig.~\ref{fig:reverseverifier} is a graphical representation of the pattern.

\vspace{0.5em}\noindent \textbf{Context:} 
In a software system, where blockchain is one of the components, the off-chain components might need to use the data stored on the blockchain and the smart contracts running on the blockchain to check certain conditions.

\vspace{0.5em}\noindent \textbf{Problem:} 
Some domains use very large and mature (or even legacy) systems, which comply with existing standards. For such domain, how to integrate the existing complex systems with blockchain in an non-intrusive approach without changing the core of the existing systems?

\vspace{0.5em}\noindent \textbf{Forces:} The problem requires to balance the following forces:
\begin{itemize}
  \item \textit{Connectivity.} Integrating blockchain into an existing system to leverage the unique properties of blockchain, as discussed in Section~\ref{sec:properties}.
  \item \textit{Simplicity.} Introduce minimal changes to the existing system.
\end{itemize}

\vspace{0.5em}\noindent \textbf{Solution:} 
The unique ID of the transactions or blocks on blockchain is a piece of data that can be easily integrated into the existing systems so that they can refer to specific transactions as having taken place or address data permanently stored in specific blocks. Reverse Oracle is a component sitting between the blockchain and the other components in the system. Reverse Oracle is mainly reading data from the blockchain, and inserting the reference (ID of transaction) of this data into other components in the system. Oracle, however, is mainly writing data into the blockchain from other components. Validation of such data is implemented by smart contracts running on blockchain. Any off-chain component is required to query the blockchain through using the ID of the referenced data. More technical details of implementing the pattern could be found in ~\cite{2020-Muehlberger-BPM-BC}.

\vspace{0.5em}\noindent \textbf{Consequences:} 

Benefits:
\begin{itemize}
  \item \textit{Connectivity.} The blockchain is integrated into an existing system with minimal effort.
\end{itemize}

Drawbacks: 
\begin{itemize}
   \item \textit{Non-intrusive.} It's not always possible to use blockchain in a non-intrusive way depending on the extensibility of the existing systems. Writing to and reading from the blockchain might need changing the existing system so that they can securely access the blockchain network.
\end{itemize}


\vspace{0.5em}\noindent \textbf{Related patterns:} 
\begin{itemize}
    \item \textit{Centralized Oracle} (Section~\ref{sec:centralOracle}) and \textit{Decentralized Oracles} (Section~\ref{sec:decentralOracle}) can be used when the information flows from external to blockchain.
\end{itemize}

\vspace{0.5em}\noindent \textbf{Known uses:}
\begin{itemize}
  \item \textit{Identitii}\footnote{\url{https://identitii.com/}} provides a solution to enrich the payments in banking systems with documents and tamper-proof attributes stored on a private blockchain. Identitii invents the concept of identity token stored on a blockchain. Every payment is associated with an identity token, which is used to exchange enriched information about a payment. The identity token is exchanged between the banks through being embedded into existing SWIFT protocol messages and can be verified against the copy in the blockchain. 
  \item \textit{Chaintrace}\footnote{\url{https://chainflux.com/wine-traceability-using-blockchain/}} records all the information about a wine, such as source, location, volume of the ingredients on blockchain. Chaintrace can be connected to external supply chain system and injects relevant information about a bottle of wine for cross verification.
\end{itemize}

\section{Data Management Patterns}
\label{sec:datamanagement}

Due to the unique properties and limitations of blockchain, the main architectural consideration for a blockchain-based software application is to decide what data and executable code (smart contract) should be kept on-chain, and what should be kept off-chain. Two factors need particular attention, namely performance and privacy. Performance highly depends on the type of deployment of the blockchain. For example, a consortium blockchain~\cite{Sherry:ICSA2017} can be configured to achieve much better performance than a public blockchain. This section discusses four data management patterns that manage data on and off blockchain.


\subsection{ \textbf{Pattern 5: Tokenisation}}
\label{sec:tokenisation}

\noindent \textbf{Summary:} Using tokens on blockchain to represent fungible goods for easier distribution. 

\vspace{0.5em}\noindent \textbf{Context:} 
The concept of tokenisation has emerged centuries ago with the first currency systems. Tokenisation is a means to reduce risk in handling high value financial instruments by replacing them with equivalents, for example, the tokens used in casino. Tokens can represent a wide range of goods which are transferable and fungible, like shares, or tickets. Blockchain is a suitable technique for asset management because of its immutability and transparency. 

\vspace{0.5em}\noindent \textbf{Problem:} 

How to have a representative of assets to avoid repetition and decrease risk?

\vspace{0.5em}\noindent \textbf{Forces:} The problem requires to balance the following forces:
\begin{itemize}
  \item \textit{Risk.} Handling fungible financial assets with high value is risky, e.g., lost 
  assets cannot be replaced.
  \item \textit{Repetition.} An asset should be represented by only one token as the authoritative source. 
\end{itemize}

\vspace{0.5em}\noindent \textbf{Solution:} 
Tokenisation is a process starting from an asset (\eg, money) is locked under a custody (\eg, a bank), and gets represented in the cryptographic world through a token. The ownership of the digital token matches the ownership of the corresponding asset. The reverse process can take place by which the user redeems the token to recover the value which is sitting within the bank. 

Blockchain provides a trustworthy platform to realise tokenisation. There are different ways to implement tokenisation using blockchain. A token on blockchain is the authoritative source of the physical asset. Naive tokens on a blockchain (\eg, BTC on Bitcoin, ETC on Ethereum) can be used to formulate a system where the tokens represent monetary value or other physical assets. The token is generally used to track title over the physical assets. Transactions on blockchain record the verifiable title transfer from one user to another. However, using the native token on blockchain for tokenisation is limited because it can only implement the title transfer of the physical assets, with limited conditions checking. A more flexible way is to define a data structure in a smart contract to represent physical assets. By using smart contracts, some conditions can be implemented and associated with the ownership transfer.

\vspace{0.5em}\noindent \textbf{Consequences:} 

Benefits:
\begin{itemize}
  \item \textit{Risk.} Tokenisation reduces risk in handling high value financial instruments by replacing them with equivalents.
  \item \textit{Repetition.} Blockchain and smart contracts provide a trustworthy infrastructure to provide authorised tokens for the corresponding assets.
\end{itemize}

Drawbacks: 
\begin{itemize}
   \item \textit{Integrity.} Integrity of the tokens is guaranteed by the blockchain infrastructure. But the authenticity of the corresponding physical/digital asset is not guaranteed automatically.  
   \item \textit{Standardisation.} 24\% of the existing financial smart contracts on Ethereum uses this tokenisation pattern~\cite{designPatterns1}. Given the popularity of this pattern, ERC20\footnote{\url{https://eips.ethereum.org/EIPS/eip-20}} (and ERC777\footnote{\url{https://eips.ethereum.org/EIPS/eip-777}} as an advanced version) has been proposed as a fungible token standard that describes the functions and events that a token smart contract has to implement. The new proposed fungible tokens should follow the standard. 
   \item \textit{Legal processes for ownership.} A token on a blockchain is not necessarily the authoritative source of information about the ownership of a physical asset. The owner of an asset may be entitled to sell the asset without being required to create a transaction on the blockchain. Also, legal processes such as court orders and bankruptcy proceedings can change the ownership of physical assets without any associated transaction being recorded on the blockchain.
\end{itemize}


\vspace{0.5em}\noindent \textbf{Related patterns:} 
\begin{itemize}
    \item \textit{Off-Chain Data Storage} (Section~\ref{sec:hash}) can be used to add a hash of a digital asset as an ID on blockchain.
\end{itemize}

\vspace{0.5em}\noindent \textbf{Known uses:}
\begin{itemize}
  \item \textit{Digix}\footnote{\url{https://digix.global/}} uses tokens to track the ownership of gold as a physical property.
  \item \textit{Elevated Returns}\footnote{\url{https://www.elevatedreturns.com/}} is an asset management firm that uses tokenisation to manage ownership on real estates. 
  \item \textit{CargoX}\footnote{\url{https://cargox.io/}} creates a smart token to replace their bill of lading. The ownership of goods are claimed by using the smart token.
\end{itemize}



\subsection{ \textbf{Pattern 6: Off-Chain Data Storage}}
\label{sec:hash}

\begin{figure}[t]
\begin{center}
\includegraphics[scale=0.9]{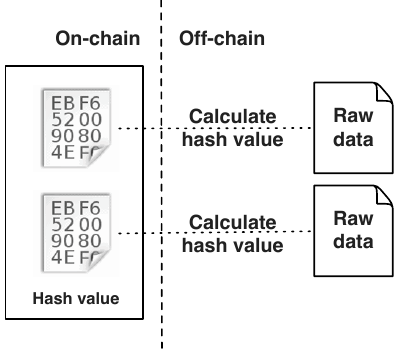}
\caption{Off-chain Data Storage Pattern}\label{fig:hashintegrity}
\end{center}
\end{figure}

\vspace{0.5em}\noindent \textbf{Summary:} Use hashing to ensure the integrity of arbitrarily large datasets which may not fit directly on the blockchain. Fig.~\ref{fig:hashintegrity} is a graphical representation of the pattern solution.

\vspace{0.5em}\noindent \textbf{Context:} Some applications consider using the blockchain to guarantee the integrity of large amounts of data.

\vspace{0.5em}\noindent \textbf{Problem:} The blockchain, due to its full replication across all participants of the blockchain network, has limited storage capacity. Storing large amounts of data within a transaction may be impossible due to the limited size of the blocks of the blockchain (for example, the gas limit on Ethereum
). Data cannot take advantage of the immutability or integrity guarantees without being stored on the blockchain. How to store data of arbitrary size and take advantage of the immutability and integrity guarantees provided by the blockchain?

\vspace{0.5em}\noindent \textbf{Forces:} The problem requires to balance the following forces:
\begin{itemize}
  \item \textit{Integrity.} Applications leverage blockchain to achieve data integrity.
  \item \textit{Scalability.} Blockchain provides limited scalability because every bit of data is replicated across all nodes, where it is kept permanently.
  \item \textit{Cost.} If a public blockchain is used, storing data on blockchain costs real money, although the cost is a one-time cost to write the data. This is in contrast to traditional distributed data storage, like cloud, which charge based on the amount of allocated storage space over time. A piece of data can be stored on blockchain through being embedded into a transaction, or as a variable of smart contract or as a log event. Embedding data into a transaction is the cheapest way, while storing data in a contract is more efficient to enable manipulation, but can be less flexible due to the potential constraints of the smart contract languages on the value types and length~\cite{Sherry:ICSA2017}. Different blockchain has different cost model for storing data.

  \item \textit{Size.} There are limits of transaction size or block size. For example, on Bitcoin blockchain, The default Bitcoin client only relayed $\mathit{OP\_RETURN}$ transactions up to 80 bytes, which was reduced to 40 bytes in 2014\footnote{\url{https://github.com/bitcoin/bitcoin/pull/3737}}. Ethereum has a block gas limit that restricts the amount of gas which all transactions in a block are allowed to use. 
\end{itemize}

\vspace{0.5em}\noindent \textbf{Solution:} The blockchain can be used as a general-purpose replicated database, as transactions logged in the blockchain can include arbitrary data on some blockchain platforms. For data of big size (essentially data that is bigger than its hash value), rather than storing the raw data directly on blockchain, a representation of the data with smaller size can be stored on blockchain with other small sized metadata about the data (\eg, a URI pointing to it). The solution is to store a hash value (also called digest) of the raw data on chain. The value is generated by a hash function, \eg one from the SHA2~\cite{SHA2} family, which maps data of arbitrary size to data of fixed size. Hash function is a one-way function which is easy to compute, but hard to invert given the output of a random input. If even one bit of the data changes, its corresponding hash value would change as well. The hash value is used for ensuring the integrity of the raw data stored off-chain, and the transaction on blockchain that includes the hash value guarantees the integrity of the hash value as well as the original raw data from which the hash was derived.

\vspace{0.5em}\noindent \textbf{Consequences:} 

Benefits:
\begin{itemize}
  \item \textit{Integrity.} Blockchain guarantees the integrity of the hash value that represents the raw data. The integrity of the raw data can be checked using the on-chain hash value.
  \item \textit{Cost.} If a public blockchain is used, blockchain is utilized at a lower cost (fixed cost as the size of the hash value is fixed) for integrity of data with arbitrary size.
\end{itemize}

Drawbacks: 
\begin{itemize}
  \item \textit{Integrity.} The raw data is stored off-chain, where the off-chain data store might not be as secure as blockchain. The raw data may be changed without authorization. This change will be detected thanks to the hash of the original data stored on the blockchain. However, without additional measures, it will neither be possible to recover the original data nor to prevent the change from happening in the first place.
  \item \textit{Availability.} Since the raw data is stored off-chain, it may be deleted or lost. Only its hash value remains permanently on the blockchain.
  \item \textit{Data sharing.} The on-chain data can be shared through using blockchain platforms. Extra communication mechanisms and storage platforms are required for data sharing off-chain.
\end{itemize}


\vspace{0.5em}\noindent \textbf{Related patterns:} 
\begin{itemize}
    \item \textit{Tokenisation} (Section~\ref{sec:tokenisation}) can be used to add a authoritative representative of an asset on blockchain.
    \item \textit{Legal and Smart Contract Pair} (Section~\ref{sec:legal}) is enabled by \textit{Off-Chain Data Storage}.
\end{itemize}

\vspace{0.5em}\noindent \textbf{Known uses:}
\begin{itemize}
  \item \textit{Proof-of-Existence (POEX.IO\footnote{\url{https://poex.io/}})} allows entering an SHA-256 cryptographic hash of a document into the Bitcoin blockchain as a ``proof-of-existence'' of the document at a certain time. The hash value guarantees the data integrity of the document.
  \item \textit{Chainy\footnote{\url{https://chainy.info/}}} is a smart contract running on Ethereum blockchain. Chainy stores a short link to an off-chain file and its corresponding hash value in one place. 
\end{itemize}

\subsection{ \textbf{Pattern 7: State channel}}
\label{sec:statechannel}

\begin{figure}[t]
\begin{center}
\includegraphics[scale=0.9]{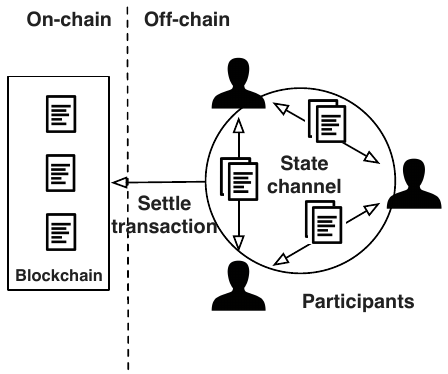}
\caption{State Channel Pattern}\label{fig:statechannel}
\end{center}
\end{figure}

\noindent \textbf{Summary:} Micro-payments transactions are too expensive to be performed on-chain because the required transaction fee might be higher than the monetary value associated with the transaction assuming a public blockchain is used. Thus, micro-payments should be exchanged off-chain while periodically recording settlements for larger amounts on chain. Such a payment channel can be generalized for arbitrary state updates for more general purposes other than monetary value. Fig.~\ref{fig:statechannel} is a graphical representation of the pattern. 

\vspace{0.5em}\noindent \textbf{Context:} 
Micro-payments are payments that can be as small as a few cents, \eg, payment of a very small amount of money to a WiFi hot-spot for every 10 kilobytes of data usage. Blockchain has potential to be used for such financial transactions with tiny monetary value. The question is if it is necessary and cost effective to store all the micro-payment transactions on blockchain.

\vspace{0.5em}\noindent \textbf{Problem:} 
The decentralized design of blockchain has limited performance. Transactions can take several minutes or even one hour (for Bitcoin blockchain) to be \emph{committed} on the blockchain~\cite{SRDS2017}. Due to the long commit time and high transaction fees on a public blockchain (where fees are largely independent of the transacted amount), it is often infeasible to store every micro-payment transaction on the blockchain network.
During a peak in demand, the average fee per transaction raised to the equivalent of US\$55\footnote{ \url{https://bitinfocharts.com/comparison/bitcoin-transactionfees.html} for 22 Dec 2017; accessed on 11/03/2021.} on Bitcoin. On-chain transactions are suitable for transactions with medium to large monetary value, relative to the transaction fee.

\vspace{0.5em}\noindent \textbf{Forces:} The problem requires to balance the following forces:
\begin{itemize}
  \item \textit{Trustworthiness.} Payment transactions on the blockchain are trusted.
  \item \textit{Latency.} Blockchain transactions may take a long time to be committed while users expect micro-payments to happen instantaneously.
  \item \textit{Scalability.} Blockchain has limited scalability because every bit of data is replicated across all nodes, and kept permanently.
  \item \textit{Cost.} Storing data on a public blockchain costs real money. The transaction fee of individual micro-payment transaction might be higher than the monetary value associated with the micro-payment transaction.
\end{itemize}

\vspace{0.5em}\noindent \textbf{Solution:} 
Storing every micro-payment transaction on blockchain is infeasible in certain contexts due to the small monetary value associated with it. Thus, a solution is to establish a payment channel between two participants, with a deposit from one or both sides of the participants locked up as security in a smart contract for the lifetime of the payment channel. The payment channel keeps the intermediate states of the micro-payment off-chain, and only stores the finalized payment on chain. The frequency of transaction settlement depends on the use case, and agreement between the two sides. For example, in scenarios around utilities, internet service providers or electricity companies can establish payment channel with their consumers for an agreed billing period, for example, a month. As the consumer uses data or energy daily, the intermediate state is stored in the channel until the end of the month, when the channel is closed to finalize the payment of that month. A network of micro-payment channels can be built where the transactions transferring small values occur off-chain. The individual transactions take place entirely off the blockchain and exclusively between the participants, across multiple hops where needed. Only the final transaction that settles the payment for a given channel or set of channels is submitted to the blockchain. The technologies used to implement state channel are normally specific to blockchain platform. For example, Lightning network\footnote{\url{https://lightning.network/}} on the Bitcoin blockchain is a proposed implementation of Hashed Timelock Contracts (HTLCs)\footnote{\url{https://en.bitcoin.it/wiki/Hashed_Timelock_Contracts}} with bi-directional payment channels which allows secure payments across multiple peer-to-peer channels. A HTLC is a type of payments that use the features of Script, like \textit{hashlocks} and \textit{timelocks}, to require that the receiver of a payment acknowledges receiving the payment prior to a deadline by generating cryptographic proof.

\vspace{0.5em}\noindent \textbf{Consequences:} 

Benefits:
\begin{itemize}
  \item \textit{Speed.} Without involving blockchain for every transfer, the off-chain transactions can be settled without waiting for the blockchain network to process the transaction, generate a new block with the transaction and reach consensus, and the desired number of confirmation blocks.
  \item \textit{Throughput.} The number of off-chain transactions that can be processed is not limited by the configuration of blockchain, such as the block size, block interval, gas limit, etc., and thus a much higher throughput can be achieved than for on-chain transactions.
  \item \textit{Privacy.} Other than the final settlement transaction, the individual off-chain transactions do not show up in the public ledger, thus, the detail of these intermediate off-chain transactions is not publicly visible.
  \item \textit{Cost.} If a public blockchain is used, only the final settlement transaction costs transaction fee to be included in the blockchain. Direct individual off-chain transactions do not cost any money. Multi-hop transactions may cost small transaction fees, which are typically charged as a percentage of the transacted amount.
\end{itemize}

Drawbacks: 
\begin{itemize}
   \item \textit{Trustworthiness.} The individual off-chain micro-payment transactions might not be as trustworthy as the on-chain transactions because the micro-payment transactions are not stored in an immutable data store. The intermediate state of payment channels might be lost after the payment channels are closed. 
   \item \textit{Liquidity.} To establish a payment channel, money from one or both sides of the channel needs to be locked up in a smart contract for the lifetime of the payment channel. The liquidity of the channel participants is thereby reduced.
   \item \textit{Wallet.} A new wallet or extension to the existing wallet is needed to support the micro-payment protocol.
\end{itemize}


\vspace{0.5em}\noindent \textbf{Related patterns:} 
\begin{itemize}
    \item \textit{Off-Chain Data Storage} (Section~\ref{sec:hash}) is another mechanism that reduces data on blockchain. 
\end{itemize}

\vspace{0.5em}\noindent \textbf{Known uses:}
\begin{itemize}
  \item The \textit{Lightning network} uses an off-chain protocol to enable micro-payments of Bitcoin and several other crypto-currencies. 

  \item The \textit{Raiden network}\footnote{\label{raiden}\url{https://raiden.network/}} on the Ethereum blockchain is a similar solution as lightning network. The basic idea is to avoid the consensus bottleneck by leveraging a network of off-chain payment channels that allow to securely transfer monetary value. Smart contracts are used to deposit value into the payment channels.
  
  
  \item \textit{State channel} on Ethereum\footnote{\url{http://www.jeffcoleman.ca/state-channels/}} and \textit{Gnosis Go}\footnote{\url{https://forum.gnosis.pm/t/how-offchain-trading-will-work/63}} offer a more generalized form of state channels that support exchanging state for general-purpose applications. 

\end{itemize}


\subsection{ \textbf{Pattern 8: Legal and Smart Contract Pair}}
\label{sec:legal}

\begin{figure}[t]
\begin{center}
\includegraphics[scale=0.9]{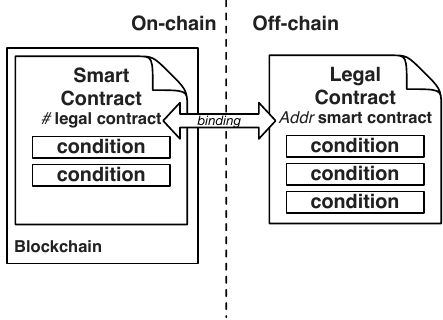}
\caption{Legal and Smart Contract Pair Pattern}\label{fig:legal}
\end{center}
\end{figure}

\noindent \textbf{Summary:} A bidirectional binding is established between a legal agreement and the corresponding smart contract. Fig.~\ref{fig:legal} is a graphical representation of the pattern.

\vspace{0.5em}\noindent \textbf{Context:} 
The legal industry is becoming digitized, for example, using digital signatures has become a valid way to sign legal agreements. The Ricardian contract~\cite{Ricardian} was developed in the mid 1990s to interpret legal contracts digitally without losing the value of the legal prose. Digital legal agreements need to be executed and enforced.

\vspace{0.5em}\noindent \textbf{Problem:} 
An independent trustworthy execution platform trusted by all the involved participants is needed to execute the digital legal agreement. How to bind a legal agreement to the corresponding smart contract on a trusted execution environment to ensure a 1-to-1 mapping?

\vspace{0.5em}\noindent \textbf{Forces:} The problem requires to balance the following forces:
\begin{itemize}
  \item \textit{Authoritative source.} A 1-to-1 mapping is required between a legal contract and its corresponding smart contract to make the smart contract as the authoritative source of the legal contract.
  \item \textit{Secure storage.} A secure and trustworthy data storage is required to keep the legal agreement.
  \item \textit{Secure execution.} A trustworthy computational platform is required to execute digital agreements to enforce certain conditions as defined in a legal contract. 
\end{itemize}

\vspace{0.5em}\noindent \textbf{Solution:} 

Blockchain can be an ideal trusted platform to run digital legal agreements, which are bound with the corresponding on-chain smart contracts. The smart contract implements conditions defined in the legal agreement. When deployed, there is a variable to store the hash value of the legal agreement, but is initially a blank value. The address of the smart contract is included in the legal agreement, and then the hash of the legal agreement is calculated and added to the contract variable. 
By binding a physical agreement with a smart contract, the bridge between the off-chain physical agreement and the on-chain smart contract is established. The two directional binding makes sure that the legal agreement and smart contract have a 1-to-1 mapping.

The smart contract digitizes the conditions defined within the legal agreement. Thus, these conditions can be checked and enforced automatically by the smart contract. However, not all the legal terms can be easily digitalized. The smart contract can also enable automated regulatory compliance checking in terms of the required information and process. However, the capability of compliance checking might be limited due to the constraints of smart contract programming language.

\vspace{0.5em}\noindent \textbf{Consequences:} 

Benefits:
\begin{itemize}
  \item \textit{Automation.} Some of the conditions defined in the legal contract, for example, a conditional payment, can be automatically enforced by blockchain. 
  \item \textit{Audit trail.} Blockchain permanently records all historical transactions related to the legal contract and the contract itself. This immutable data enables auditing at anytime in future.
  \item \textit{Clarification.} Encoding legal terms expressed in natural language into smart contracts will require to give them a clear interpretation.
\end{itemize}

Drawbacks: 
\begin{itemize}
   \item \textit{Expressiveness.} Smart contracts are written in programming languages. The smart contract languages might have limited expressiveness to express contractual terms of arbitrary complexity. The capability of regulatory compliance checking also depends on the expressiveness of the smart contracts. A regulation may regulate the process, for example, what should or should not be done by whom at what stage. 
   \item \textit{Enforceability.} If a public blockchain is used, there is no central administering authority to decide a dispute, or perform the enforcement of a court judgment. 
   \item \textit{Interpretability.} There might be different ways to interpret a certain legal term and to encode them in the smart contract. Ambiguity of natural language makes it a challenge to accurately implement a certain legal term in a way that is agreed upon by all the involved participants.
\end{itemize}



\vspace{0.5em}\noindent \textbf{Related patterns:} 
\begin{itemize}
    \item \textit{Legal and Smart Contract Pair} is enabled by \textit{Off-Chain Data Storage} (Section~\ref{sec:hash}) .
\end{itemize}

\vspace{0.5em}\noindent \textbf{Known uses:}
\begin{itemize}
  \item \textit{OpenLaw}\footnote{\url{http://openlaw.io/}} is a platform that allows lawyers to make legally binding and self-executable agreements on the Ethereum blockchain. The legal agreement templates are stored on a decentralized data storage, IPFS\footnote{\url{https://ipfs.io/}}. Users can create customized contracts for specific uses.
  \item \textit{Smart Contract Template} proposed by Barclays\footnote{\url{https://www.barclays.co.uk/}} uses legal document templates to facilitate smart contracts running on Corda\footref{corda} blockchain platform~\cite{barrclays1,barrclays2}. 
  \item Specific proposals for the representation of machine-interpretable legal terms have been explored in KWM's project on digital and analog (\textit{DnA}) contracts\footnote{\url{https://github.com/KingandWoodMallesonsAU/Project-DnA}} and in the \textit{Accord Project}\footnote{\url{https://www.accordproject.org/}}.
\end{itemize}

\section{Security Patterns}
\label{sec:securitypatterns}

This section discusses four security patterns that mainly concern the security aspect~\cite{securitypatterns} of the blockchain-based applications.

\subsection{ \textbf{Pattern 9: On-Chain Encryption}}
\label{sec:encrypted}

\begin{figure}[t]
\begin{center}
\includegraphics[scale=0.9]{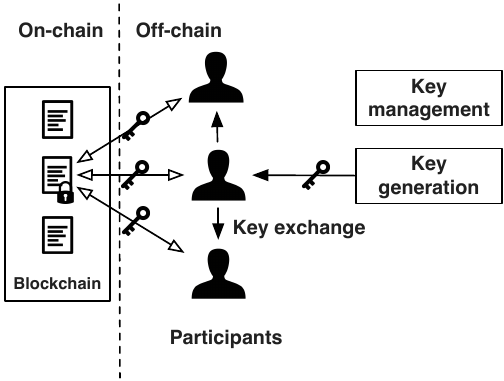}
\caption{Encrypting On-chain Data Pattern}\label{fig:encrypteddata}
\end{center}
\end{figure}

\noindent \textbf{Summary:} Ensure confidentiality of the data stored on blockchain by encrypting it. Fig.~\ref{fig:encrypteddata} is a graphical representation of the pattern.

\vspace{0.5em}\noindent \textbf{Context:} 
For some applications on blockchain, there might be commercially critical data that should be only accessible to the involved participants. An example would be a special discount price offered by a service provider to a subset of its users. Such information should not be accessible to the other users who do not get the discount. 

\vspace{0.5em}\noindent \textbf{Problem:} 
The lack of data privacy is one of the main limitations of blockchain. All the information on blockchain is publicly available to the participants of the blockchain network. There is no privileged user within the blockchain network, no matter the blockchain is public, consortium or private. On a public blockchain, new participants can join the blockchain network freely and access all the information recorded on blockchain. Any confidential data on public blockchain is exposed to the public.

\vspace{0.5em}\noindent \textbf{Forces:} The problem requires to balance the following forces:
\begin{itemize}
  \item \textit{Transparency.} Every participant within a blockchain network is able to access all the historical transactions on blockchain, which is required to enable them to validate previous transactions. The transactions on a public blockchain are also accessible to everyone with access to the internet, simply using tools like a blockchain explorer such as Etherscan\footnote{\label{etherscan}\url{http://etherscan.io}}. 
  
  \item \textit{Lack of confidentiality.} Since all the information on blockchain is publicly available to everyone in the network, commercially sensitive data meant to be kept confidential should not be stored on blockchain, at least not in plain form.
\end{itemize}

\vspace{0.5em}\noindent \textbf{Solution:} 
To preserve the privacy of the involved participants, symmetric or asymmetric encryption can be used to encrypt data before inserting the data into blockchain. One possible design for sharing encrypted data among multiple participants is as follows. First, one of the involved participants creates a secret key for encrypting data and distributes it during an initial key exchange. When one of the participants needs to add a new data item to the blockchain, they first symmetrically encrypt it using the secret key. Only the participants allowed to access the transaction have the secret key and can decrypt the information. 


\vspace{0.5em}\noindent \textbf{Consequences:} 

Benefits:
\begin{itemize}
  \item \textit{Confidentiality.} Using encryption, the publicly accessible information on blockchain is encrypted, so that is useless to anyone who does not hold the secret key. 
\end{itemize}

Drawbacks: 
\begin{itemize}
  \item \textit{Compromised key.} Both symmetric and asymmetric encryption require off-chain key management. If key management is not done properly, it can lead to compromise and disclosure of private or secret keys. If the required private key or secret key is compromised, the encryption mechanism does not guarantee the confidentiality nor the integrity of the data. 
  \item \textit{Access revocation.} Revoking read access is a challenge after the encrypted data has been published to the blockchain. The encrypted data on blockchain is immutable. Thus, as long as the participant keeps the secret key used to encrypt the data, it has access to the encrypted data forever. 
  \item \textit{Immutable data.} Even if stored in encrypted form, the critical data will remain in the blockchain forever. In addition to the risk of key compromise, the encrypted data may be subject to brute force decryption attacks at any time in the future, or breakthroughs in technology like quantum computing might render current encryption technologies ineffective. So even if the data is considered to be secure with a given key size when it is stored in the blockchain, this may no longer be the case in the future.
  \item \textit{Key sharing.} The encryption key needs to be shared off-chain before submitting any relevant transaction to the blockchain secretly. Although blockchain can be used as a software connector~\cite{sherry2016} to communicate data, secret keys can not be shared through blockchain because the shared key would be publicly accessible if being communicated through blockchain.
\end{itemize}


\vspace{0.5em}\noindent \textbf{Related patterns:}
\begin{itemize}
    \item \textit{Off-Chain Data Storage} (Section~\ref{sec:hash}) also provides data confidentiality because the raw data is not stored on blockchain.
\end{itemize}

\vspace{0.5em}\noindent \textbf{Known uses:}
\begin{itemize}
  \item Encrypted queries from \textit{Provable}\footref{note}. Provable is a smart contract running on Ethereum public blockchain, which provides a service to access state from external world. Provable allows smart contract developers to encrypt the parameters of their queries locally by using a public key before passing them to a smart contract. The only one who can decrypt the call parameters is Provable with the paired private key. 
  
  \item \textit{Crypto digital signature} is suggested by \textit{MLGBlockchain}\footnote{\url{https://mlgblockchain.com/crypto-signature.html}} to encrypt data and share the data between the parties who interact and transmit data through blockchain. 
  
  \item Hawk~\cite{Hawk} is a smart contract system that stores transactions as encrypted data on blockchain to retain the privacy of the transactions. The compiler of Hawk can automatically generate a cryptographic protocol for a smart contract. The involved participants interact with the blockchain following the cryptographic protocol. 
\end{itemize}


\subsection{ \textbf{Pattern 10: Multiple Authorization}}
\label{sec:authority}

\begin{figure}[t]
\begin{center}
\includegraphics[scale=0.9]{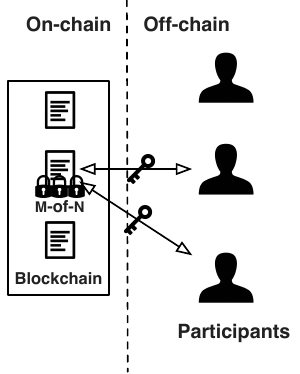}
\caption{Multiple Authorization Pattern}\label{fig:multipleauthority}
\end{center}
\end{figure}

\noindent \textbf{Summary:} A set of blockchain addresses which can authorise a transaction is pre-defined. Only a subset of the addresses is required to authorize transactions. Fig.~\ref{fig:multipleauthority} is a graphical representation of the pattern.

\vspace{0.5em}\noindent \textbf{Context:} 
In blockchain-based applications, activities might need to be authorized by multiple blockchain addresses. For example, a monetary transaction may require authorization from multiple blockchain addresses.

\vspace{0.5em}\noindent \textbf{Problem:} 
The actual addresses that authorize an activity might not be able to be decided due to the availability of the authorities. How to allow multiple authorities to dynamically authorize a transaction based on their avilability?

\vspace{0.5em}\noindent \textbf{Forces:} The problem requires to balance the following forces:
\begin{itemize}
  \item \textit{Flexibility.} The actual authorities who authorize the transaction can be from a set of pre-defined authorities. 
  \item \textit{Tolerance of compromised or lost private key} Authentication on blockchain uses digital signature. However, blockchain does not offer any mechanism to recover a lost or a compromised private key. Losing a key results in permanent loss of control over an account, and potentially smart contracts that refer to it. 
\end{itemize}

\vspace{0.5em}\noindent \textbf{Solution:} 
It would enable more dynamism if the set of blockchain addresses for authorization are not decided before the corresponding transaction being submited into the blockchain network, or the corresponding smart contract being deployed on blockchain. On the Bitcoin blockchain, a multi-signature mechanism can be used to require more than one private key to authorize a Bitcoin transaction. In Ethereum, smart contract can mimic multi-signature mechanism. 

\vspace{0.5em}\noindent \textbf{Consequences:} 

Benefits:
\begin{itemize}
  \item \textit{Flexibility.} This pattern enables flexible binding of authorities, but depends on the availability of authorities when the activity is proceeded. 
  \item \textit{Lost key tolerant.} One participant can own more than one blockchain address to reduce the risk of losing control over their smart contracts due to a lost private key. There could be a function that can update the list of allowed authorities, and the threshold of the authorization. In the case that the update function also requires threshold-based authorization, the list of the update addresses can be also updated through authorization from at least the minimum number of addresses.
\end{itemize}

Drawbacks: 
\begin{itemize}
  \item \textit{Pre-defined authorities.} Although the pattern enables flexible binding, all the possible authorities still need to be known in advance of any decision or update. 
  \item \textit{Lost key.} At least 
  a minimum number of private keys should be safely kept to avoid losing control.
  \item \textit{Cost of dynamism.} If a public blockchain is used, updating the list of authorities costs money, as does deploying the logic for multiple authorities. Besides, it costs more to store multiple addresses as the possible authorities than storing only one. 
\end{itemize}


\vspace{0.5em}\noindent \textbf{Related patterns:}
\begin{itemize}
    \item \textit{Dynamic Authorization} (Section~\ref{sec:dynamic}). An off-chain secret enabled dynamic authorization pattern is used when the possible authorities are unknown beforehand. 
    \item \textit{Voting} (Section~\ref{sec:voting}) can be enabled by Multiple Authorization.
\end{itemize}

\vspace{0.5em}\noindent \textbf{Known uses:}
\begin{itemize}
  \item Multi-Signature mechanism provided by Bitcoin\footnote{\url{https://en.bitcoin.it/wiki/Multisignature}}.
  \item Multi-signature wallet, written in Solidity, running on Ethereum blochchain and is available in the Ethereum DApp browser Mist\footnote{\label{mist}\url{https://github.com/ethereum/mist}}.
\end{itemize}


\subsection{ \textbf{Pattern 11: Dynamic Authorization}}
\label{sec:dynamic}

\begin{figure}[t]
\begin{center}
\includegraphics[scale=0.9]{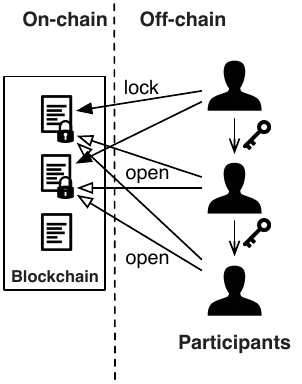}
\caption{Off-chain Secret Enabled Dynamic Authorization Pattern}\label{fig:dynamic}
\end{center}
\end{figure}

\noindent \textbf{Summary:} Using a hash created off-chain to dynamically bind authority for a transaction. Fig.~\ref{fig:dynamic} is a graphical representation of the pattern. This solution is sometimes referred to as \emph{Hashlock}.

\vspace{0.5em}\noindent \textbf{Context:} 
In blockchain-based applications, some activities need to be authorized by one or more participants that are unknown when a first transaction is submitted to blockchain.

\vspace{0.5em}\noindent \textbf{Problem:} 
Sometimes, the authority who can authorize a given activity is unknown when the corresponding smart contract is deployed, or the corresponding transaction is submitted to the blockchain. Blockchain uses digital signature for authentication and transaction authorization. Blockchain does not support dynamic binding with an address of a participant which is not defined in the respective transaction or smart contract. All accounts that can authorize a second transaction have to be defined in the first transaction before that transaction is added to the blockchain. How to allow one or more unknown authorities to dynamically authorize a transaction?

\vspace{0.5em}\noindent \textbf{Forces:} The problem requires to balance the following forces:
\begin{itemize}
  \item \textit{Dynamism.} Dynamic binding one or more unknown authorities with a second transaction representing an activity after the first transaction submitted to blockchain.
  \item \textit{Pre-defined authorities.} Using only on-chain mechanisms, all the possible authorities are required to be defined beforehand.
\end{itemize}

\vspace{0.5em}\noindent \textbf{Solution:} 
An off-chain secret can be used to enable a dynamic authorization when the participant authorizing a transaction is unknown beforehand. In the context of payment, for example, a smart contract can be used as an escrow. When the sender deposits the money to an escrow smart contract, a hash of a secret (for example, a random string, called pre-image) is submitted with the money as well. Whoever receives the secret off-chain can claim the money from the escrow smart contract by revealing the secret.
With this solution, the receiver of the money does not need to be defined beforehand in the escrow contract. This can be generalized to any transaction that needs authorization from a dynamically bound participant. Note that since the secret is revealed, it cannot be reused. One variant is to lock multiple transactions with the same secret -- by unlocking one, all of them are unlocked.

\vspace{0.5em}\noindent \textbf{Consequences:} 

Benefits:
\begin{itemize}
  \item \textit{Dynamism.} This pattern enables dynamic binding of unknown authorities after the transaction is added into the blockchain.
  
  \item \textit{Lost key tolerant.} No specific private key is required to authorize transactions.
  
  \item \textit{Routability.} This pattern has the useful property that once the secret is revealed, any other transactions secured using the same secret can also be opened. This makes it possible to create multiple transactions that are all locked by the same secret. This property is used by micro-payment channels (Section~\ref{sec:statechannel}) to enable multi-hop transfers where the money hosted by every hop and secured by a same secret can be released after the end receiver claims the money with the secret (\ie\ the secret is revealed). The secret can be exchanged through an off-chain channel to every hop.   
  
  \item \textit{Interoperability.} There is no need for a special protocol to exchange the secret. The secret can be exchanged in any ways off-chain. It provides a mechanism for other systems to trigger events on blockchain. 
\end{itemize}

Drawbacks: 
\begin{itemize}
    \item \textit{One-off secret.} The secret used in this pattern is a one-off secret. Verification of the secret is on-chain. Thus, once a secret is embedded in a transaction submitted to the blockchain, the secret is revealed. 
    \item \textit{Combination of signature and secret.} Because this pattern has the property that once the secret is revealed, any other transactions secured using the same secret can also be opened, sometimes the transaction protected by the secret should also be associated with a public key so that both a correct secret and an appropriate signature with the respective private key are required to authorize the transaction. This is applicable to the situation where a large set of authorities are known beforehand, but not all of them are allowed to authorize a certain activity/transaction. Thus, a hash secret is used to dynamically bind one or multiple authorities from the larger pre-defined set of authorities.  
    \item \textit{Lost secret.} The sender/initiator of a transaction takes the risk of losing the off-chainsecret. If the secret is lost, the transaction cannot be authorized and being proceeded anymore. In the case of money transfer, the money associated with the transaction would be locked forever if the transaction cannot be authorized properly.
\end{itemize}


\vspace{0.5em}\noindent \textbf{Related patterns:}
\begin{itemize}
    \item \textit{Multiple Authorization} (Section~\ref{sec:authority}). The multiple authorization pattern is used when all the possible authorities are known beforehand. Multiple authorization pattern is an on-chain mechanism.  
    \item \textit{Voting} (Section~\ref{sec:voting}) can be enabled by Dynamic Authorization.
    \item Dynamic Authorization enables routability for \textit{State channel} (Section~\ref{sec:statechannel}) in the financial transaction scenarios.
    
\end{itemize}

\vspace{0.5em}\noindent \textbf{Known uses:}
\begin{itemize}
  \item \textit{Raiden network}\footref{raiden} is a network of off-chain payment channels on top of Ethereum blockchain network, which enables secure value transfer. The multi-hop transfer mechanism in Raiden Network uses \emph{hashlocked} transactions to securely router payment through a middleman.
  \item In the Bitcoin ecosystem, \textit{atomic cross-chain trading}\footnote{\url{https://en.bitcoin.it/wiki/Atomic_cross-chain_trading}} allows one crytocurrency (for example, Bitcoin) to be traded for another cryptocurrency (for example, tokens on a Bitcoin sidechain) using a off-chain hash secret.
\end{itemize}

\subsection{ \textbf{Pattern 12: Embedded Permission}}
\label{sec:permission}

\begin{figure}[t]
\begin{center}
\includegraphics[scale=0.9]{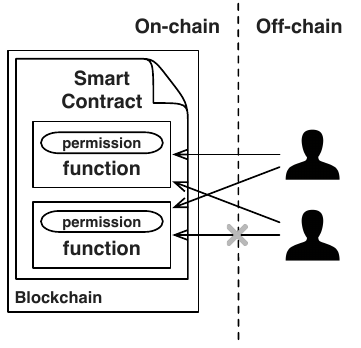}
\caption{Embedded Permission Pattern}\label{fig:permission}
\end{center}
\end{figure}

\noindent \textbf{Summary:} Smart contracts use an embedded permission control to restrict access to the invocation of the functions defined in the smart contracts. Fig.~\ref{fig:permission} is a graphical representation of the pattern.

\vspace{0.5em}\noindent \textbf{Context:} 
All the smart contracts running on blockchain can be accessed and called by all the blockchain participants and other smart contracts by default, because there are no privileged users in blockchain network. In the case of public blockchain, every one with internet access can join the network to access all the information and code stored and running on blockchain. 

\vspace{0.5em}\noindent \textbf{Problem:} 
A smart contract by default has no owner, meaning that once deployed the author of the smart contract has no special privilege on the smart contract. A permission-less function can be triggered by unauthorized users accidentally. Such a permission-less function becomes vulnerability of blockchain-based application. For example, a permission-less function which is discovered in a smart contract library used by the Parity multi-signature wallet, caused the freezing of about 500K Ethers\footnote{\url{https://paritytech.io/a-postmortem-on-the-parity-multi-sig-library-self-destruct/}}. 7\% smart contract on public Ethereum can be terminated without authority~\cite{SRDS2017}.


\vspace{0.5em}\noindent \textbf{Forces:} The problem requires to balance the following forces:
\begin{itemize}
  \item \textit{Security.} The functions defined in the smart contracts should be called only by the authorized participants. Due to the transparency of public blockchains, all the smart contracts are also publicly available to everyone connecting to the Internet. In contrast, in a conventional software system, the internal logic is normally not visible to the end uses. Interaction with the software system is either through a user interface or API, where it is possible to enforce access control policies. 
\end{itemize}

\vspace{0.5em}\noindent \textbf{Solution:} 
Adding permission control to every smart contract function to check permissions for every caller that triggers the functions defined in the smart contract based on the blockchain addresses of the caller. This can be done by checking the authorization of the caller before executing the logic of the function: unauthorized calls are rejected and the execution of the function terminated before reaching the core logic of the function. 

\vspace{0.5em}\noindent \textbf{Consequences:} 

Benefits:
\begin{itemize}
  \item \textit{Security.} Only the participants and smart contracts that are authorized by the smart contract can call the corresponding functions successfully.
  \item \textit{Secure authorization.} Authorization is implemented in smart contracts running on blockchain, which leverages the properties provided by blockchain. 
\end{itemize}

Drawbacks: 
\begin{itemize}
   \item \textit{Cost.} On a public blockchain, extra code that implements the permission control mechanism also has additional monetary cost for deployment and execution.
    \item \textit{Lack of flexibility.} Such permissions are defined in the smart contract before its deployment, therefore they are difficult to change. However, permissions may be required to be dynamic. A mechanism is needed to support dynamic granting and removal of permissions.
\end{itemize}


\vspace{0.5em}\noindent \textbf{Related patterns:}
\begin{itemize}
    \item \textit{Multiple authorization} (Section~\ref{sec:authority}) is one way of implementing embedded permission pattern.
    \item \textit{Dynamic authorization} (Section~\ref{sec:dynamic}) is another way of implementing embedded permission pattern.
\end{itemize}

\vspace{0.5em}\noindent \textbf{Known uses:}
\begin{itemize}
  \item The Mortal contract discussed in the Solidity tutorial\footnote{\url{http://solidity.readthedocs.io/en/develop/contracts.html}} restricts the permission of invoking the \textit{selfdestruct} function to the ``owner'' of the contract -- where ``owner'' is a variable defined in the contract code itself.
  \item The \textit{Restrict access} pattern suggested in the Solidity tutorial\footnote{\url{http://solidity.readthedocs.io/en/develop/common-patterns.html}} uses \textit{modifier} to restrict who can make modifications to the state of the contract or call the functions of the contract. \textit{Modifier} is a mechanism that adds a piece of code before the function to check certain conditions. 
\end{itemize}

\section{Contract Structural Patterns}
\label{sec:contractPatttern}

This section discusses five smart contracts patterns. Essentially, smart contracts are programs running on blockchain, thus some of the existing design patterns and programming principles for conventional software environments are also applicable to smart contracts. If a public blockchain is used, the structural design of the smart contract has large impact on its deployment and execution cost. The cost of deploying a smart contract depends on the size of the smart contract(s) because the code is stored on blockchain, resulting in a data storage fee that is proportional to the size of the smart contract. Thus, a structural design with more lines of compiled code costs more money. A consortium blockchain does not necessarily have tokens/currency; therefore monetary cost is typically not an issue for a consortium blockchain. However, blockchain size is still a design concern because the total size of the blockchain keeps growing as more blocks are appended to it and no block can ever be detached from it, and every participant stores a full replica of blockchain. Besides, different structural designs of smart contracts may affect performance because more or less transactions may be required.


\subsection{ \textbf{Pattern 13: Contract Registry}}
\label{sec:registry}

\begin{figure}[t]
\begin{center}
\includegraphics[scale=0.9]{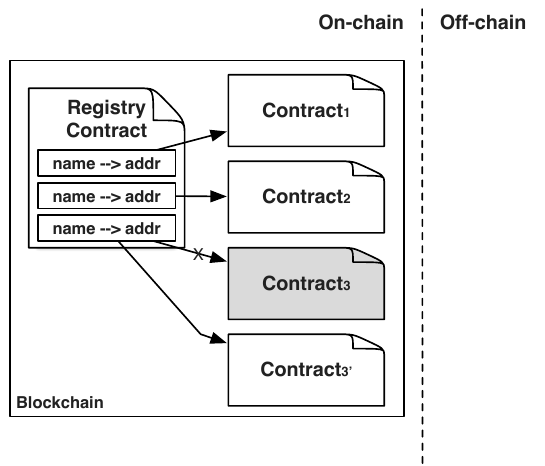}
\caption{Contract Registry Pattern}\label{fig:registry}
\end{center}
\end{figure}

\noindent \textbf{Summary:} Before invoking it, the address of the latest version of a smart contract is located by looking up its name on a contract registry. Fig.~\ref{fig:registry} is a graphical representation of the pattern.

\vspace{0.5em}\noindent \textbf{Context:}
As any software application, blockchain-based applications need to be upgraded to new versions. To do so, the on-chain functions defined in smart contracts need to be updated to fix bugs as well as to fulfill
new requirements. 

\vspace{0.5em}\noindent \textbf{Problem:} 
Smart contracts deployed on blockchain cannot be upgraded because the code of the smart contracts as a type of data, stored on blockchain is immutable. How to perform upgrades of smart contracts? 

\vspace{0.5em}\noindent \textbf{Forces:} The problem requires to balance the following forces: \begin{itemize}
  \item \textit{Immutability.} Every bit of data, including deployed smart contracts, stored on blockchain is immutable.
  \item \textit{Upgradability.} There is a fundamental need to upgrade all but short-lived applications and their smart contracts over time. 
  \item \textit{Human-readable contract identifier.} The identifier of a smart contract on blockchain platforms, like Ethereum, is hexadecimal address, which is not human-readable.

\end{itemize}

\vspace{0.5em}\noindent \textbf{Solution:} 
An on-chain registry contract is used to maintain a mapping between user-defined symbolic names and the blockchain addresses of the registered contracts. The address of the registry contract needs to be advertised off-chain. The creator of a contract can register the name and the address of the new contract to the registry contract after the new contract being deployed. The invoker of a registered contract retrieves the latest version of the new smart contract from the registry contract. The corresponding functions provided by the registered contract can be upgraded by replacing the address of the old version contract in the registry contract with the address of a new version without breaking the dependency between the upgraded smart contract and other smart contracts that depend on its functions. The address of a contract is stored as a variable in the registry contract. The value of contract variables can be updated. The registry contract can have a permission control module to maintain the writing permission. Note that all the previous values of the variable are still stored on the blockchain. 

\vspace{0.5em}\noindent \textbf{Consequences:} 

Benefits:
\begin{itemize}
  \item \textit{Human-readable contract name.} The registry contract maintains a mapping between human-readable names and the hexadecimal addresses of the smart contracts. A human readable form of smart contract names is desired, for example, to be exposed to the user interface. A human readable name is also useful for developers.
  \item \textit{Constant contract name.} The smart contract associated with a registered name can be updated without changing its name. This way dependencies relying on the name of the smart contract do not get broken.
  \item \textit{Upgradability.} The smart contract associated with a registered name could be replaced by a new version without breaking the dependencies based on the human-readable name. 
  \item \textit{Version control.} Version control can be integrated in the registry contract as well to allow a look-up based on the name and version of a smart contract. Old versions of a smart contract that are no longer needed should be terminated. 
\end{itemize}

Drawbacks: 
\begin{itemize}
  \item \textit{Upgradability.} Upgradability is still limited if the functions defined in the smart contract are called by other contracts. Although the implementation of the function can be upgraded, the interface (that is function signature) cannot be modified without breaking the link to dependent smart contracts. Similar methods as for API / service interface management need to be implemented, \eg through versioning and depreciation flags.
  \item \textit{Cost.} There is an additional cost to maintain a registry that contains the mapping between the contract names and their addresses. Furthermore, all the inter-contract function calls require a registry look-up to find the latest version of the smart contract to be invoked.
\end{itemize}


\vspace{0.5em}\noindent \textbf{Related patterns:}
\begin{itemize}
    \item \textit{Embedded permission} (Section~\ref{sec:permission}) can be used to define writing permission. 
    \item \textit{Data contract} (Section~\ref{sec:separate}) and this pattern can work together to further improve upgradability of smart contracts. 
\end{itemize}

\vspace{0.5em}\noindent \textbf{Known uses:}
\begin{itemize}
  \item ENS\footnote{\url{https://ens.domains}} is a name service on Ethereum blockchain, which is implemented as smart contracts. ENS maintains a mapping between  both smart contracts on-chain and resources off-chain and simple, human-readable names. ENS can be viewed as a contract registry built on Ethereum blockchain, which is accessible to everyone. A blockchain-based application can also maintain a separate registry contract for the application. 
  \item \textit{KairosFuture}\footnote{\url{https://www.kairosfuture.com/publications/reports/the-land-registry-in-the-block-chain-testbed/}} experimented using blockchain smart contract as a replacement for land registry. All the signed contracts are recorded by a registry contract on the blockchain.
  \item \textit{Kaleido}\footnote{\url{https://www.kaleido.io/blockchain-blog/smart-contract-management-solution-how-it-works-why-you-need-it}} proposed a enterprise-level smart contract management solution. The core function of their management solution is the contract registry that is accessable to every participants in Kaleido network. 
\end{itemize}

\subsection{ \textbf{Pattern 14: Data Contract}}
\label{sec:separate}

\begin{figure}[t]
\begin{center}
\includegraphics[scale=0.9]{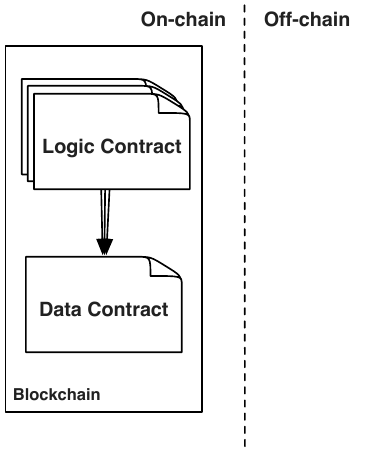}
\caption{Data Contract Pattern}\label{fig:separated}
\end{center}
\end{figure}

\noindent \textbf{Summary:} Store data in a separate smart contract. Fig.~\ref{fig:separated} is a graphical representation of the pattern.

\vspace{0.5em}\noindent \textbf{Context:} 
The need to upgrade a blockchain-based application over time is ultimately necessary, so as the smart contracts used by the application. In general, logic and data change at different times and with different frequencies. 
There are different ways to store a data on blockchain, as discussed in \emph{Off-chain Data Storage} pattern (Section~\ref{sec:hash}).

\vspace{0.5em}\noindent \textbf{Problem:} 
Storing data on blockchain is expensive and there is a limitation on the amount of data and amount of computation a transaction can contain. In the context of upgrading smart contracts, the upgrading transactions might contain a large data storage for copying the data from the old version of the smart contract to the new version of the smart contract. Porting data to a new version might even require multiple transactions, \eg when the block gas limit on Ethereum prevents an overly complex data migration transaction.

\vspace{0.5em}\noindent \textbf{Forces:} The problem requires to balance the following forces:
\begin{itemize}
  \item \textit{Coupling.} Smart contracts can live forever on blockchain if not being explicitly terminated. If a smart contract is deactivated in this way, the data stored in the smart contract cannot be accessed through the smart contract functions any more -- although it can still be accessed with some effort, \eg for provenance or audit purposes.
  \item \textit{Upgradability.} The need to upgrade the application and the smart contracts supporting the application over time is ultimately necessary for many applications. 
  \item \textit{Cost.} If a public blockchain is used, storing data on blockchain costs money. Thus copying data from an old version of a smart contract to a new version should be avoid or minimized. 
\end{itemize}

\vspace{0.5em}\noindent \textbf{Solution:} 
To avoid moving data during upgrades of smart contracts, the data store is isolated from the rest of the code. 
In the context of blockchain, data could be separately stored in different smart contracts to enable isolation. Depending on the circumstances of the application, how large of a data store it needs and whether the data structure is expected to change often, the data store could use a strict definition or a loosely typed flat store. The more generic and flexible data structure can be used by all the other logic smart contracts and is unlikely to require changes. One example of a generic data structure is a mapping to store SHA-256 key and value pairs.

\vspace{0.5em}\noindent \textbf{Consequences:} 

Benefits:
\begin{itemize}
  \item \textit{Upgradability.} By separating data from the rest of the code, the logic of the application is able to be upgraded without affecting the data contract.
  \item \textit{Cost.} Since the data is separated from the rest of the code, there is no cost for migrating data when the application is upgraded.
  \item \textit{Generality.} If the data can be cleanly separated and generalized, there would be an additional benefit: the generic data contract can be used by all related logic smart contracts. 
\end{itemize}

Drawbacks: 
\begin{itemize}
   \item \textit{Cost.} If a public blockchain is used, storing a piece of data in a generic data structure costs more money than a strictly defined data structure. For example, as mentioned earlier, a generic data structure maintains a mapping between SHA-256 key and value pairs, but a more strictly defined data structure can be of smaller size, \eg not requiring the key to be stored. Querying the data is also less straightforward. This is the cost of a generalized solution.  
\end{itemize}


\vspace{0.5em}\noindent \textbf{Related patterns:}
\begin{itemize}
    \item \textit{Contract registry} (Section~\ref{sec:registry}) and Data Contract can work together to further improve upgradability of smart contracts.
\end{itemize}


\vspace{0.5em}\noindent \textbf{Known uses:}
\begin{itemize}
  \item \textit{Chronobank}\footnote{\url{https://chronobank.io/}} is a blockchain project that tokenizes labour and provides a market for professionals to trade their labour time with businesses. It uses a smart contract with a generic data structure as the data store used by all the other logic smart contracts.
  \item \textit{Colony}\footnote{\url{https://colony.io/}}, a platform for open organizations running on Ethereum. Similar to Chronobank, Colony has a data contract with a generic data structure. 
\end{itemize}


\subsection{ \textbf{Pattern 15: Factory Contract}}
\label{sec:factory}

\begin{figure}[t]
\begin{center}
\includegraphics[scale=0.9]{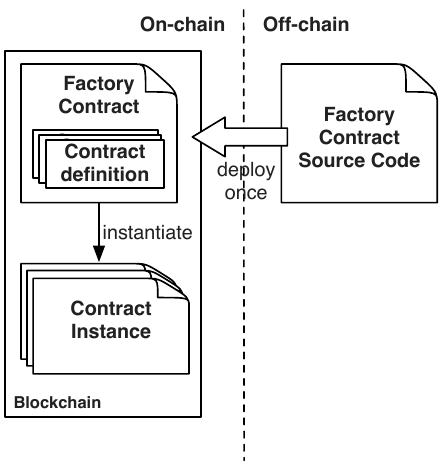}
\caption{Factory Contract Pattern}\label{fig:factory}
\end{center}
\end{figure}

\noindent \textbf{Summary:} An on-chain template contract is used as a factory that generates contract instances from the template. Fig.~\ref{fig:factory} is a graphical representation of the pattern.

\vspace{0.5em}\noindent \textbf{Context:} 
Applications based on blockchain might need to use multiple instances of a standard contract with customization. Each contract instance is created by instantiating a contract template. For example, in a business process management system, each of the business process instances might be represented by a smart contract being generated from a contract template representing the business process model~\cite{Weber:BPM2016}. The template can be stored off-chain in a code repository, or on-chain, within its own smart contract.

\vspace{0.5em}\noindent \textbf{Problem:} 
Keeping the contract template off-chain cannot guarantee consistency between different smart contract instances created from the same template because the source code of the template can be independently modified. 

\vspace{0.5em}\noindent \textbf{Forces:} The problem requires to balance the following forces:

\begin{itemize}
  \item \textit{Dependency management.} Storing the source code of smart contract off-chain in a code repository introduces the issue of integrating more systems into the blockchain-based application.
  \item \textit{Secure code sharing.} The source code smart contract should be stored in a secure storage. 
  \item \textit{Deployment.} If a public code repository, like Github, is used to store the source code of a smart contract, a component is needed to implement the function of deploying smart contract on blockchain, otherwise, the end users need to understand how to deploy smart contracts.
\end{itemize}

\vspace{0.5em}\noindent \textbf{Solution:} 
Smart contracts are created from a contract factory deployed on blockchain. The factory contract is deployed once from the off-chain source code. The factory may contain the definition of multiple smart contracts. Smart contract instances are generated by passing parameters to the contract factory to instantiate customized smart contract instances. Factory contract is analogous to a \textit{Class} in an object-oriented programming language. 
Every transaction that generates a smart contract instance essentially instantiates an object of the factory contract class. This contract instance (the object) will maintain its own properties independently of the other instances but with a structure consistent with its original template. 

\vspace{0.5em}\noindent \textbf{Consequences:} 

Benefits:
\begin{itemize}
  \item \textit{Security.} Keeping the factory contract on-chain guarantees the consistency of the contract definition. Blockchain provides a secure platform to share code of smart contracts. As opposed to a traditional code repository, changes of code deployed on a smart contract can be strictly limited or prohibited.
  \item \textit{Efficiency.} If the contract definition is kept on-chain in a factory contract, smart contract instances are generated by calling a function defined in the factory contract. 
\end{itemize}

Drawbacks: 
\begin{itemize}
  \item \textit{Deployment cost.} If a public blockchain is used, using factory contract requires extra cost to deploy the factory contract. 
  \item \textit{Function call cost.} If a public blockchain is used, creating a new smart contract instance requires extra cost to call a function defined in the factory contract. 
\end{itemize}


\vspace{0.5em}\noindent \textbf{Related patterns:}
\begin{itemize}
    \item \textit{Contract registry} (Section~\ref{sec:registry}) can be used to store the addresses of all the smart contract instances generated from a factory contract. The factory and instance registry can be implemented in the same contract, although that limits upgradability.
\end{itemize}

\vspace{0.5em}\noindent \textbf{Known uses:}
\begin{itemize}
  \item A tutorial from Ethereum developers\footnote{\url{https://ethereumdev.io/manage-several-contracts-with-factories/}} about how to create a contract factory from which smart contract instances can be created.
  \item Factory pattern has been applied in a real-world blockchain-based health care application~\cite{factorypattern}.
  \item The business process management system in an academic work~\cite{Weber:BPM2016} uses a contract factory to generate process instances.
\end{itemize}


\subsection{ \textbf{Pattern 16: Incentive Execution}}
\label{sec:incentiveexecution}

\begin{figure}[t]
\begin{center}
\includegraphics[scale=0.9]{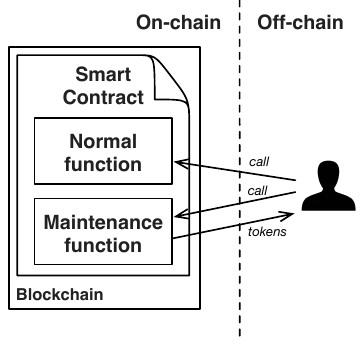}
\caption{Incentive Execution Pattern}\label{fig:incentiveexecution}
\end{center}
\end{figure}

\noindent \textbf{Summary:} Reward is provided to the caller of the contract function for invoking the execution. Fig.~\ref{fig:incentiveexecution} is a graphical representation of the pattern.

\vspace{0.5em}\noindent \textbf{Context:} 
Smart contracts are event-driven programs, which cannot execute autonomously. All the functions defined in a smart contract need to be triggered either by a transaction from external account or another smart contract to execute. Other than the functions that provide regular services to users, some functions need to run asynchronously from regular user interaction, for example, to clean up the expired records, or make dividend payouts etc. Such accessorial functions usually involve a time, after which the function should start.

\vspace{0.5em}\noindent \textbf{Problem:} 
Users of a smart contract have no direct benefit from calling the accessorial functions. If a public blockchain is used, executing these functions causes extra monetary cost. Some accessorial functions are expensive to execute. How to make sure the the accessorial functions are invoked?

\vspace{0.5em}\noindent \textbf{Forces:} The problem requires to balance the following forces:
\begin{itemize}
  \item \textit{Completeness.} The regular services provided by a smart contract are supported by some accessorial functions.
  \item \textit{Cost.} Execution of accessorial functions causes extra costs from the users.
\end{itemize}

\vspace{0.5em}\noindent \textbf{Solution:} 
Reward the caller of a function defined in a smart contract for invoking the execution, for example, sending back a percentage of payout to the caller to reimburse the (gas) execution cost. 

\vspace{0.5em}\noindent \textbf{Consequences:} 

Benefits:
\begin{itemize}
  \item \textit{Completeness.} The execution of the accessorial functions helps to complete the regular services provided by the smart contract.
  \item \textit{Cost.} The users, who spends extra to execute the accessorial functions, are compensated by the reward associated with the execution.
\end{itemize}

Drawbacks: 
\begin{itemize}
  \item \textit{Execution.} Execution cannot be guaranteed even with incentive. Thus, another option is to embed the logic of accessorial functions into other regular functions that users have to call to use the services.
\end{itemize}


\vspace{0.5em}\noindent \textbf{Related patterns:} 
\begin{itemize}
    \item Both \textit{Security Deposit} (Section~\ref{sec:securitydeposit}) and Incentive Execution provide incentive mechanisms.
\end{itemize}

\vspace{0.5em}\noindent \textbf{Known uses:}
\begin{itemize}
  \item \textit{Ethereum alarm clock}\footnote{\label{alarm}\url{http://www.ethereum-alarm-clock.com/}} is a service provided by a smart contract running on Ethereum. It facilitates scheduling function calls for a specified block in the future and provides incentive for users to execute the scheduled function. 
\end{itemize}


\subsection{ \textbf{Pattern 17: Security Deposit}} 
\label{sec:securitydeposit}

\begin{figure}[t]
\begin{center}
\includegraphics[scale=0.9]{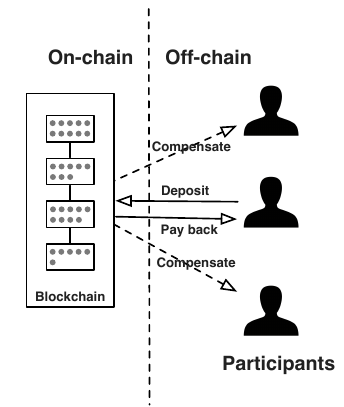}
\caption{Security Deposit Pattern}\label{fig:securitydeposit}
\end{center}
\end{figure}

\noindent \textbf{Summary:} A user put aside a certain amount of money, which will be paid back to the user for her honesty or given to the other parties to compensate them for the dishonesty of the user. Fig.~\ref{fig:securitydeposit} is a graphical representation of the pattern.

\vspace{0.5em}\noindent \textbf{Context:} 
In a decentralized environment like blockchain, trust in the blockchain-based application is achieved from the interactions between participants within the network. A blockchain-based application rely on all the users rather than relying on trusted third-party organisations to facilitate transactions.

\vspace{0.5em}\noindent \textbf{Problem:} 
The \textit{equal rights} property of blockchain allows every participant the same ability to access and manipulate the blockchain and the decentralized applications running on blockchain. How can a certain participant proves to others that he/she behaves honestly?

\vspace{0.5em}\noindent \textbf{Forces:} The problem requires to balance the following forces:
\begin{itemize}
  \item \textit{Security.} The security of a decentralized application relies on the behaviour of all the participants of the application. 
  \item \textit{Incentive.} In a decentralized application, participants can be incentivised to behave honestly.
\end{itemize}

\vspace{0.5em}\noindent \textbf{Solution:} 
Initially, participant is required to put aside amount of tokens, which will be paid back to the participant if she behave honestly, otherwise the deposit is given to the other parties to compensate them for their loss. Such security deposit is recorded on blockchain, which is publicly available to other parties. Security deposit is a way to reduce the risk of participants in the network misbehave by temporarily sacrificing some of her stake.

\vspace{0.5em}\noindent \textbf{Consequences:} 

Benefits:
\begin{itemize}
  \item \textit{Security.} The security of the application is obtained through security deposit because the deposit will be paid back to the participant only if the participant behaves honestly. 
\end{itemize}

Drawbacks: 
\begin{itemize}
    \item \textit{Access.} The security deposit is normally larger than the potential profit the participant gain from her/his dishonesty. If the required deposit is large, it restricts some participants to access the application. 
    \item \textit{Liquidity.} Similarly as above, the liquidity of the application participants is reduced due to the deposit.
\end{itemize}

\vspace{0.5em}\noindent \textbf{Related patterns:} 
\begin{itemize}
    \item Both \textit{Incentive Execution} (Section~\ref{sec:incentiveexecution}) and Security Deposit provide incentive mechanisms.
\end{itemize}

\vspace{0.5em}\noindent \textbf{Known uses:}
\begin{itemize}
  \item The notion of ``deposits'' has already been used in \textit{Bitcoin}'s contract\footnote{\url{https://en.bitcoin.it/wiki/Contract\#Example_1:_Providing_a_deposit}}, where deposit is bought by a party with no reputation as a proof of her trust. 
  \item \textit{Ethereum alarm clock}\footref{alarm} enables scheduling of transactions for delayed execution in the future. Before the execution window, there is a claim window when the request may be claimed by a participant for execution. To claim a request, the participant is required to put down a deposit. If the claimer fulfils their commitment to execute the request, the deposit is returned to them, otherwise, the deposit is given to someone else that executes the request as an additional reward. 
 \end{itemize}

\section{User Interaction Patterns}
\label{sec:deploymentPatttern}

This section illustrates two patterns regarding user interacting with smart contracts.


\subsection{\textbf{Pattern 18: DApp}}
\label{sec:DApp}

\begin{figure}[t]
\begin{center}
\includegraphics[scale=0.9]{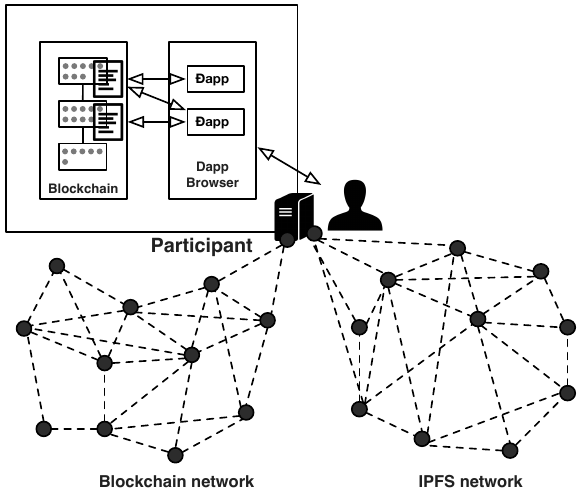}
\caption{DApp Pattern}\label{fig:Dapp}
\end{center}
\end{figure}

\noindent \textbf{Summary:} Decentralized applications (DApps) are applications running on P2P network rather than a single computer. DApps are blockchain-based web applications that allow users to interact with smart contracts deployed on blockchain. Fig.~\ref{fig:Dapp} is a graphical representation of the pattern.

\vspace{0.5em}\noindent \textbf{Context:} 
Users interact with smart contracts through sending transactions to invoke the functions defined in smart contracts. In order for the users to understand how to interact with a smart contract, the source code of the smart contract is open source and visible to users. Second, the application binary interface (ABI) of the smart contract is also publicly accessible so that users can send transactions to it. 

\vspace{0.5em}\noindent \textbf{Problem:} 
Users need a strong technical understanding of blockchain and smart contract to be able to generate transactions calling smart contracts. Such process is error-prone and with a bad user experience~\cite{patternICSA2018}. How to call a smart contract in a trustless environment?

\vspace{0.5em}\noindent \textbf{Forces:} 
The problem requires to balance the following forces:
\begin{itemize}
  \item \textit{Learning Curve.} Users need to understand the functionality of a smart contract before being able to interact with it. To understand the input required and the output produced by a smart contract, users need to read the documentation or the source code of the smart contract. 
  \item \textit{Convenience.} Manually generating transactions to interact with an smart contract is an error-prone process despite the understanding of the smart contract. Depending on whether many transactions need to be sent, it may be worth to invest into automating the transaction submission process.
  \item \textit{Trust.} The user needs to trust the provider of such automated transaction submission process.  
\end{itemize}

\vspace{0.5em}\noindent \textbf{Solution:} 
DApp provides a front-end user interface for users to easily interact with smart contracts. The frond end uses the same technology as conventional web applications to render web pages. The difference is that DApps can be hosted on a decentralized storage, like IPFS, and are rendered by DApp browsers, like Ethereum Mist\footref{mist} or a plug-in to a web application browser, like MetaMask\footnote{\url{https://metamask.io/}}. Both the smart contract on blockchain and the DApp is deployed on a local node.  The transactions calling smart contracts are generated by DApps, and are presented to the users for further verification before being sent to the blockchain.

\vspace{0.5em}\noindent \textbf{Consequences:} 

Benefits:
\begin{itemize}
  \item \textit{Convenience.} Users interact with smart contract through a front-end provided by the DApp provider instead of reverse engineering how to submit transactions from the source code of the smart contract. The user experience of using the front-end of a DApp are much better as it requires less technical understanding of the smart contract. Assuming the DApp is correct, it is also less error-prone than manually generating transactions to interact with smart contracts.
\end{itemize}

Drawbacks: 
\begin{itemize}
   \item \textit{Trust.} Using DApps imposes trust to the DApp provider because the transactions are generated by the DApp on behalf of the user. However, the impact of the execution is not explicit without understanding the smart contracts underneath, especially if the source code of the contract or ABI is not published. 
   \item \textit{Learning Curve.} Basic technical knowledge regarding transactions and \textit{gas}  are still required by users to use DApps and verify the transactions generated by the DApps.
\end{itemize}

\vspace{0.5em}\noindent \textbf{Related patterns:}
\begin{itemize}
    \item \textit{Semi-DApp} (Section~\ref{sec:SemiDApp}) can be browsed using a conventional web browser without any DApp plugin.
\end{itemize}

\vspace{0.5em}\noindent \textbf{Known uses:}
\begin{itemize}
  \item \textit{State of the DApps}\footnote{\url{https://www.stateofthedapps.com/}} provides a directory of the DApps on Ethereum blockchain. There are 1800+ DApps with different levels of maturity registered to the directory.  
  \textit{DappRadar}\footnote{\url{https://dappradar.com/}} also provides a directory of DApps, not only on Ethereum blockchain but also included DApps deployed on EOS\footnote{\url{https://eos.io/}} and TRON\footnote{\url{https://tron.network/index?lng=en}}. 
  \item \textit{Remix}\footnote{\url{https://remix.ethereum.org}} is an open source smart contract compiler that allow user to write and interact with Solidity contracts from the browser.
\end{itemize}


\subsection{\textbf{Pattern 19: Semi-DApp}}
\label{sec:SemiDApp}

\begin{figure}[t]
\begin{center}
\includegraphics[scale=0.9]{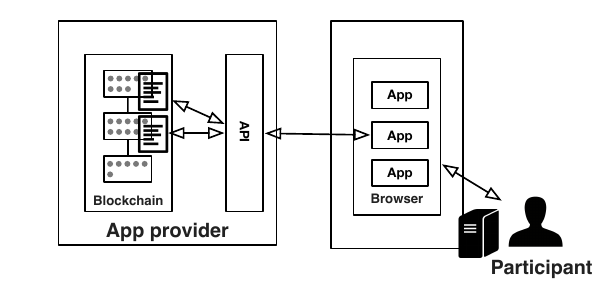}
\caption{Hybrid-DApp Pattern}\label{fig:Dapp2}
\end{center}
\end{figure}

\noindent \textbf{Summary:} Semi-DApp provider offers a web application, which can be browsed using a conventional web browser without any DApp plugin (like MetaMask). Fig.~\ref{fig:Dapp2} is a graphical representation of the pattern.

\vspace{0.5em}\noindent \textbf{Context:} 
Users interact with smart contracts through sending transactions to invoke the functions defined in smart contracts. In order for the users to interact with a smart contract, the source code of the smart contract is open source and visible to users. Second, the application binary interface (ABI) of the smart contract is also publicly accessible. 

\vspace{0.5em}\noindent \textbf{Problem:} 
Users need a strong technical understanding of blockchain and smart contract to be able to generate transactions and interact with smart contracts. Such process is error-prone, with a bad user experience~\cite{patternICSA2018}. Even with a front-end that assists the user to interact with smart contracts, some basic technical knowledge regarding transactions and \textit{gas} is still required by users to use DApps and verify the transactions generated by the DApps (Section~\ref{sec:DApp}). It is not feasible for non-technical user to learn and understand the source code of smart contract to use it.

\vspace{0.5em}\noindent \textbf{Forces:} 
The problem requires to balance the following forces:
\begin{itemize}
  \item \textit{Learning Curve.} Users need some basic technical knowledge of smart contracts in order to use DApps. 
  \item \textit{User Experience.} Due to the decentralization nature of DApps, the front-end of most DApps is quite simple, and exposes too many technical details of the underlying smart contract, which not all users are capable to grasp.
\end{itemize}

\vspace{0.5em}\noindent \textbf{Solution:} 
Semi-DApp provider offers a standard web application for users to easily interact with smart contracts. The web application keeps the technical details of the corresponding smart contracts opaque to the users. The website communicates with the backend through RESTful API calls
Thus, the backend is responsible for interacting with the smart contracts on behalf of the user who is not able to inspect nor validate the transactions being sent to the blockchain. The transaction ID is normally returned to the user for the user to check the transaction on blockchain using blockchain explorer, like EtherScan\footref{etherscan}.

\vspace{0.5em}\noindent \textbf{Consequences:} 

Benefits:
\begin{itemize}
  \item \textit{Convenience.} Users interact with smart contracts through a web application provided by the Semi-DApp provider. The user experience of using the front-end of such a DApp is as same as using the front-end of a conventional web application. This solution offers the maximum convenience and the most gentle learning curve. 
\end{itemize}

Drawbacks: 
\begin{itemize}
   \item \textit{Trust.} Users need to completely trust the Semi-DApp provider who is responsible for managing their private keys. One example is a Japan-based Bitcoin exchange, Mt. Gox\footnote{\url{https://en.wikipedia.org/wiki/Mt._Gox}}. Launched in 2010,  Mt. Gox used to be the largest Bitcoin exchange in the world, which handles over 70\% of all Bitcoin transactions worldwide. In June 2011, Mt. Gox announced that approximately 850,000 Bitcoins (an amount valued at more than \$450 million at the time) belonging to customers and the company disappeared, most likely due to a compromised internal computer. 
\end{itemize}

\vspace{0.5em}\noindent \textbf{Related patterns:} 
\begin{itemize}
    \item \textit{DApp} (Section~\ref{sec:DApp}) allows users to interact with smart contracts.
\end{itemize}

\vspace{0.5em}\noindent \textbf{Known uses:}
\begin{itemize}
  \item Cryptocurrency exchanges are common examples of this patterns as they always interact with the blockchain on behalf of the users. \textit{Kraken}\footnote{\url{https://www.kraken.com/}} is a San Francisco-based Bitcoin exchange. 
  \item \textit{Binance}\footnote{\url{https://www.binance.com/}} is a cryptocurrency exchange founded in China. 
\end{itemize}

\section{Related Work}
\label{sec:relatedwork}

To document the reusable solutions for blockchain-based application design, blockchain patterns have been summarized~\cite{Dilum:migrationpattern,Lu:patterns,2020-Muehlberger-BPM-BC,Eberhardt2017,patternICSA2018,securityblockchainpattern}.
Some of them are generic and can be applied for general purposes~\cite{2020-Muehlberger-BPM-BC,Eberhardt2017,patternICSA2018}. For example, five patterns are proposed for blockchain-based applications in~\cite{Eberhardt2017}, which focus on what data and computation should be on-chain and what should be kept off-chain. Other documented patterns apply to specific use cases~\cite{Dilum:migrationpattern,Lu:patterns}. 
There are also design patterns for writing smart contracts from both academia and industry\footnote{\url{https://consensys.github.io/smart-contract-best-practices/}}. The binary code of a smart contract deployed on a public blockchain is publicly accessible, and for many, the code is also open-source. Empirical analysis has been conducted using smart contracts from different public blockchain platforms to identify the common programming patterns~\cite{designPatterns1,Zdun:patterns,Zdun:patterns2}. Existing patterns from conventional programming languages have been also applied to smart contract programming, for example, four existing object-oriented software patterns were applied to smart contract programming in the context of a blockchain-based health case application~\cite{factorypattern}.
 
Compared with the above existing work, our paper covers system-level design patterns about interaction between blockchain and other components within a big software system, data management patterns, security patterns, structural patterns for smart contracts and user interaction patterns. Some structural patterns are new and some are modifications of the existing design patterns. More importantly, we provide use cases from the real world with each of the patterns. There is some overlap between the existing works and our paper. For example the \emph{Proxy} pattern from~\cite{factorypattern} is a more generic pattern compared with our \emph{Off-chain data storage} pattern. The \emph{Off-chain signatures} pattern from~\cite{Eberhardt2017} is similar to our \emph{State channel} pattern. The \emph{Authorization} pattern from ~\cite{designPatterns1} is similar to our \emph{Embedded permission} pattern. \emph{Self-Confirmed transactions} pattern from ~\cite{patternICSA2018} is similar to our \emph{DApps} pattern and \emph{Delegated transactions pattern} is similar to our \emph{Semi-DApps} patter. 


\section{Conclusions}
\label{sec:conclusions}

We view the blockchain as a fundamental building block of large-scale decentralized software systems. For effective use of blockchain to this end, patterns are needed that show how to make good use of the blockchain in the design of systems and applications. In this paper, we propose a pattern collection for blockchain-based applications. Our pattern collection includes four patterns about interaction between blockchain and the external world, four data management patterns, four security patterns, five contract structural patterns and two user interaction patterns. The pattern collection provides an architectural guidance for developers to build applications on blockchain. Some patterns are designed specifically for blockchain-based applications considering the unique properties of blockchain. Others are variants of existing software patterns applied to smart contracts. 

\section*{Acknowledge}

We want to thank Tim Wellhausen, the shepherd for our EuroPLoP2018 patterns paper, and the anonymous reviewers of our paper for their helpful comments.

\bibliographystyle{splncs04}
\bibliography{bibliography}

\end{document}